\documentclass[pra,twocolumn,showpacs,floatfix,aps]{revtex4}
\usepackage{amsmath}
\usepackage{makeidx}
\usepackage{amssymb}
\usepackage{graphicx}

\usepackage[usenames]{color}
\usepackage[normalem]{ulem}

\usepackage{hyperref}
\usepackage[T1]{fontenc} 

\begin{document}

\title{Vortex-bright solitons in a spin-orbit coupled spin-$1$ condensate}

\author{Sandeep Gautam\footnote{sandeepgautam24@gmail.com}}
\author{S. K. Adhikari\footnote{adhikari44@yahoo.com, 
        URL  http://www.ift.unesp.br/users/adhikari}}
\affiliation{Instituto de F\'{\i}sica Te\'orica, Universidade Estadual
             Paulista - UNESP, \\ 01.140-070 S\~ao Paulo, S\~ao Paulo, Brazil}
      

\date{\today}
\begin{abstract}
We study the vortex-bright solitons in a quasi-two-dimensional spin-orbit-coupled 
(SO-coupled) hyperfine spin-1 three-component Bose-Einstein condensate (BEC) using 
variational method and numerical solution of a mean-field model. The ground state 
of these vortex-bright solitons is radially symmetric for  weak ferromagnetic 
and polar interactions. For a sufficiently strong ferromagnetic interaction,
we observe the emergence of an asymmetric vortex-bright soliton as the ground 
state. We also numerically investigate  stable moving solitons and binary collision 
between them. The present mean-field model is not Galilean invariant, and we 
use a Galilean-transformed model for generating the moving solitons. At low 
velocities, the head-on collision between two {\em in-phase} solitons results either
in collapse or fusion of the soliton pair. On the other hand, in head-on
collision, the two {\em out-of-phase} solitons strongly repel each other and 
trace back their trajectories before the actual collision. 
At low velocities, in a collision with an impact parameter, the {\em out-of-phase} solitons get 
deflected from their original trajectory like two rigid classical disks. 
These {\em out-of-phase solitons} behave like classical disks, 
and their collision dynamics is governed by classical  laws of motion. 
However, at large velocities two SO-coupled spinor solitons, irrespective of
phase difference, can pass through each 
other in a head-on collision like two quantum solitons. 
  
\end{abstract}
\pacs{03.75.Mn, 03.75.Hh, 67.85.Bc, 67.85.Fg}

\maketitle

\section{Introduction}
\label{Sec-I}
A self-reinforcing solitary wave which preserves its shape while traversing
at a constant speed is known as a bright soliton. The origin of the bright solitons 
is due to a cancellation of the effects produced by non-linear and dispersive terms 
in the Hamiltonian. Solitons have been studied in a wide range of systems ranging 
from water waves, non-linear optics \cite{Kivshar}, ultracold quantum gases 
including spinor Bose-Einstein condensates (BECs) 
\cite{Inouye,li, rb, Perez-Garcia,Ieda}, {etc}.
In this paper, we study the two-dimensional (2D) vortex-bright solitons in 
spin-orbit (SO) coupled three-component spin-1 spinor Bose-Einstein condensates. 
The SO coupling is
the coupling between the spin of the atom and its center of mass motion. In neutral
atoms, the SO coupling is absent  \cite{stringari}. Nevertheless, neutral atoms can be subjected
to the SO coupling by creating a non-Abelian gauge potential by suitably 
modifying the atom-light interaction \cite{Dalibard}. The SO coupling with
equal strengths of Rashba \cite{Rashba} and Dresselhaus \cite{Dresselhaus} terms 
was first engineered in a landmark experiment with a BEC of $^{87}$Rb by dressing 
two of its internal 
spin states from within the ground electronic manifold ($5S_{1/2}, F = 1$) with a 
pair of lasers \cite{Lin}. In recent years, a variety of experimental studies
have been done on SO-coupled Bose-Einstein condensates \cite{Aidelsburger}. 
Solitonic structures have been theoretically investigated in SO-coupled quasi-one-dimensional
(quasi-1D)
\cite{rela} and quasi-2D pseudospin-1/2 condensates \cite{Xu,Sakaguchi}.
Bright solitons have also been theoretically studied in SO-coupled
quasi-1D spin-1 \cite{Liu,Gautam-3} and spin-2 condensates  \cite{Gautam-4}.

In this paper, we study the stable stationary and moving vortex-bright solitons in a  quasi-2D \cite{Salasnich} 
SO-coupled spin-1 condensate using the mean-field Gross-Pitaevskii (GP) {equations} \cite{Ohmi}. 
We observe that for small strengths of SO coupling which we employ in the paper, 
the ground state vortex-bright soliton of an SO-coupled polar and weakly-ferromagnetic  
spin-1 condensate is an axisymmetric 
vortex-bright soliton of type $(- 1,0, + 1)$ with zero magnetization, where 
the numbers in the parenthesis are the phase-winding numbers (angular momenta) \cite{Mizushima} associated with 
the spin components $m_f=+1,0,1$. An anti-vortex in component $m_f=+1$ is associated with an overlapping 
vortex of opposite circulation in $m_f=-1$ component. Besides this, we have also identified a stationary  
excited axisymmetric vortex-bright soliton of type $(0,+1,+ 2)$. 
The spin texture of this excited state vortex-bright soliton shows that it is
a {\em coreless Anderson-Toulose} vortex \cite{Mizushima}. 
For condensates with stronger ferromagnetic interaction, the ground state is 
an asymmetric vortex-bright soliton with an anti-vortex of unit charge in the spin 
component $m_f=+1$ associated with a vortex of opposite circulation in the 
$m_f=-1$ component. In this case the vortex and anti-vortex are separated from 
each other, and the separation can occur along any arbitrary direction, which will get spontaneously chosen
in an experiment, in two-dimensional plane. However, the condensate collapses for very strong ferromagnetic 
interaction and no vortex-bright soliton can be formed.

The 2D vortex-bright solitons were first suggested and studied in the pseudospin-1/2 two-component  
spin-1 BEC \cite{Sakaguchi}, which is an approximation over the present  three-component model of spin-1 BEC.
In general, the implementation of  SO interaction in the three-component spin-1 BEC is more 
complicated  
than the same in the two-component pseudospin-1/2 BEC from both theoretical \cite{gtm} and experimental \cite{campbell} 
point of view. The present study  goes beyond that previous investigation \cite{Sakaguchi}.
It provides a more intuitive understanding of the role of SO coupling in generating the 
solitons, viz. Figs. \ref{fig0}(a)-(b), in addition to  a critical study of statics and interaction dynamics of the 
2D solitons. Although these solitons behave as true solitons in frontal collision at high velocities, at low velocities, depending 
on the relative phase, they may repel and bounce back like in the collision of two rigid elastic disks or may transfer all atoms to one soliton to form a soliton molecule. Only the collision of two 1D analytic solitons is truly elastic at all velocities.

  Besides stationary vortex-bright
  solitons, we have also investigated the stable moving vortex-bright soliton of the
SO-coupled spin-1 condensate. As the present mean-field model does not possess
 Galelian invariance, the moving solitons are calculated with the Galelian-transformed model \cite{rela,Sakaguchi,Liu,Gautam-3}.
We find that  the structure of the moving vortex-bright soliton
is a function of both the magnitude and the direction of velocity, which can result
in different density distributions for vortex-bright solitons moving along different directions. 
At low velocities, the collision of two vortex-bright solitons with a phase difference of $\pi$ is 
elastic. The two solitons repel  and avoid each other and rebound from the center of collision without ever forming an 
overlapped profile. The collision dynamics is demonstrated to obey classical  laws of motion.
If  the same initial guess is used for the right and the left moving solitons
in the numerical simulation of the stationary state, the solitons acquire a  phase difference of $\pi$.  If this phase difference 
is removed before the numerical simulation of the colliding solitons,  then after the collision between the two slow moving 
vortex-bright solitons, all the atoms end up being captured by one of the solitons.
Similarly, in the collision of two normal BEC solitons at sufficiently 
low velocities, the two colliding solitons lose their identity and  form a stable overlapping 
profile called a soliton molecule \cite{Luis,Nguyen}. At large velocities, the two vortex-bright solitons 
undergo quasi-elastic collision with the two solitons crossing each other 
irrespective of the phase difference.

The paper is organized as follows. In Sec. \ref{Sec-IIA}, we describe the mean-field
coupled Gross-Pitaevskii (GP) equations with Rashba SO coupling used to study the
vortex-bright solitons in a spin-1 condensate. This is followed by a variational
analysis of the stationary axisymmetric vortex-bright solitons in Sec. \ref{Sec-IIB}.
In Sec. \ref{Sec-III}, we provide the details of the numerical method used to solve
the coupled GP equations with SO coupling. We 
discuss the numerical results for axisymmetric vortex-bright solitons in Sec. \ref{Sec-IVA},
asymmetric solitons in Sec. \ref{Sec-IVB}, stability of the  solitons in
Sec. \ref{Sec-IVC}, and moving solitons and collisions between solitons in Sec. \ref{Sec-IVD}.
Finally, in Sec. V, we give a summary of our findings.


\section{Spin-Orbit coupled   BEC vortex-bright soliton}
\label{Sec-II} 

\subsection{Mean-field equations}
\label{Sec-IIA}
For the study of a quasi-2D vortex-bright soliton, we consider a spin-1 spinor BEC under a 
harmonic trap $m\omega_z^2 z^2/2$ in the $z$ direction and free in the $x-y$ 
plane. After integrating out the $z$ coordinate, the single particle Hamiltonian 
of the condensate with Rashba \cite{Rashba} SO coupling in such a  quasi-2D trap 
is \cite{H_zhai}
\begin{equation}
H_0 = \frac{p_x^2+p_y^2}{2m}  + \gamma p_x \Sigma_x+\gamma p_y \Sigma_y,
\label{sph} 
\end{equation}
where $p_x = -i\hbar\partial/\partial x$ and $p_y = -i\hbar\partial/\partial y$ 
are the momentum operators along $x$ and $y$ axes, respectively,  
and $\Sigma_x$ and $\Sigma_y$ are the irreducible representations of the $x$ 
and $y$ components of the spin matrix, respectively,
\begin{eqnarray}
\Sigma_x=\frac{1}{\sqrt 2} \begin{pmatrix}
0 & 1 & 0 \\
1 & 0  & 1\\
0 & 1 & 0
\end{pmatrix}, \quad  \Sigma_y=\frac{1}{\sqrt 2 i} \begin{pmatrix}
0 & 1 & 0 \\
-1 & 0  & 1\\
0 & -1 & 0
\end{pmatrix},
\end{eqnarray}
and $\gamma$ is the strength of SO coupling. In the mean-field approximation, 
the SO-coupled quasi-2D spin-1 BEC is described by the following set of three 
coupled two-dimensional GP equations, written here in dimensionless form, 
for different spin components 
$m_f=\pm 1,0$ \cite{Ohmi,Kawaguchi}
 \begin{align}
i\frac{\partial \psi_{\pm 1}(\mathbf r)}{\partial t} &=
 {\cal H}\psi_{\pm 1}(\mathbf r) 
\pm   c^{}_1F_z\psi_{\pm 1}(\mathbf r) 
+ \frac{c^{}_1}{\sqrt{2}} F_{\mp}\psi_0(\mathbf r)\nonumber\\
&-\frac{i\gamma}{\sqrt{2}}\left(
  \frac{\partial\psi_0}{\partial x}\mp i\frac{\partial\psi_0}{\partial y}\right),
 \label{gps-1}\\
i \frac{\partial \psi_0(\mathbf r)}{\partial t} &=
{\cal H}\psi_0(\mathbf r)  
+ \frac{c_1}{\sqrt 2} [F_{-}\psi_{-1}(\mathbf r) 
+F_{+}\psi_{+1}(\mathbf r)]\nonumber\\
&-\frac{i \gamma}{\sqrt{2}}\Bigg(\frac{\partial\psi_{1}}{\partial x} +i \frac{\partial\psi_{1}}{\partial y} 
  +\frac{\partial\psi_{-1}}{\partial x}-i\frac{\partial\psi_{-1}}{\partial y}\Bigg)
\label{gps-2}, 
\end{align}
where ${\bf F}\equiv\{F_x,F_y,F_z\}$ is a vector whose three components are the 
expectation values of the three spin-operators over the multicomponent 
wavefunction, and is called the spin-expectation value \cite{Kawaguchi}. Also,
\begin{align}&
F_{\pm}\equiv  F_x \pm i F_y=
\sqrt 2[\psi_{\pm 1}^*(\mathbf r)\psi_0(\mathbf r)
+\psi_0^*(\mathbf r)\psi_{\mp 1}(\mathbf r)]\label{fpmspin1}, \\
&  F_z= \rho_{+1}(\mathbf r)-\rho_{-1}(\mathbf r)\label{fzspin1},
\quad
{\cal H}= -\frac{\nabla^2}{2} +c_0 \rho,\\
&c_0 = \frac{2N \sqrt{2\pi }(a_0+2 a_2)}{3 l_0},~
c_1 = \frac{2N \sqrt{2\pi }({a_2-a_0})}{3 l_0},\label{nonlin}\\
&\nabla^2=\frac{\partial ^2}{\partial x^2} + \frac{\partial ^2}{\partial y^2},~
{\mathbf r }\equiv\{x,y\}, \label{nabla_q2d}
\end{align}
where $\rho_j=|\psi_j(\mathbf r)|^2$ with $j=\pm 1, 0$ are the component densities,
$\rho=\sum_{j}\rho_j$ is the total density, $a_0$ and $a_2$ are the $s$-wave scattering 
lengths in the total spin 0 and 2 channels, respectively, and asterisk denotes 
complex conjugate. The normalization condition satisfied by the 
component wavefunctions $\psi_j$ is $\int \sum_j \rho_j d{\bf r} =1$.   
All quantities in Eqs. (\ref{gps-1})-(\ref{nabla_q2d}) are dimensionless. 
This is achieved by writing length, density, and energy in units of 
$l_0$ $(=\sqrt{\hbar/(M\omega_z)})$, $l_0^{-2}$, and $\hbar\omega_z$, respectively. 
The energy of the system in dimensionless unit is
\begin{align}
 E &= \int_{-\infty}^{\infty} d {\bf r}\Bigg\{\frac{1}{2}\left(\sum_{j=-1}^1 \left|\nabla \psi_j\right|^2 
+  {c_0}\rho^2 + {c_1}|\mathbf F|^2\right)\nonumber \\
&-
\frac{i\gamma}{\sqrt{2}}\psi_0^*
   \left(\frac{\partial \psi_1}{\partial x} + \frac{\partial \psi_{-1}}{\partial x}\right)
  +  \frac{\gamma}{\sqrt{2}}\psi_0^*\left(\frac{\partial \psi_1}{\partial y} - \frac{\partial \psi_{-1}}{\partial y}\right)
\nonumber \\
&-\frac{i\gamma}{\sqrt{2}} 
 \left(\psi_1^*+\psi_{-1}^*\right)\frac{\partial \psi_0}{\partial x}
   -\frac{\gamma}{\sqrt{2}} \left(\psi_1^*-\psi_{-1}^*\right)\frac{\partial \psi_0}{\partial y} \Bigg\}.
  \label{energy}
\end{align}

In plane polar coordinates, ${\bf r} = (r,\phi)$, Eqs. (\ref{gps-1})-(\ref{gps-2}) are
 \begin{align}
i\frac{\partial \psi_{\pm 1}(r,\phi)}{\partial t} &=
 {\cal H}(r,\phi)\psi_{\pm 1}(r,\phi) 
\pm   c^{}_1F_z\psi_{\pm 1}(r,\phi) 
\nonumber\\
&+ \frac{c^{}_1}{\sqrt{2}} F_{\mp}\psi_0(r,\phi)-\frac{i\gamma e^{\mp i\phi}}{\sqrt{2}}\left(
  \frac{\partial\psi_0}{\partial r}\mp i\frac{\partial\psi_0}{r\partial \phi}\right),
 \label{gpsp-1}\\
i \frac{\partial \psi_0(r,\phi)}{\partial t} &=
{\cal H}(r,\phi)\psi_0(r,\phi)  
+ \frac{c_1}{\sqrt 2} [F_{-}\psi_{-1}(r,\phi) 
\nonumber\\
&+F_{+}\psi_{+1}(r,\phi)]-\frac{i \gamma} {\sqrt{2}}\Bigg[e^{i\phi}\left(\frac{\partial\psi_{1}}{\partial r} 
+i \frac{\partial\psi_{1}}{r\partial \phi}\right)\nonumber\\
& +e^{-i\phi}\left(\frac{\partial\psi_{-1}}{\partial r}-i\frac{\partial\psi_{-1}}{r \partial \phi}\right)\Bigg]
\label{gpsp-2}. 
\end{align}
The coupled Eqs. (\ref{gpsp-1})-(\ref{gpsp-2}) in polar coordinates are instructive to
understand the underlying symmetries of the system.

\subsection{Vortex-bright  soliton}
\label{Sec-IIB}
This study revealed two types of stationary quasi-2D low-energy axisymmetric 
vortex-bright solitons in an SO-coupled spin-1 BEC for an attractive 
(negative) $c_0$ and for $c_1\ge c_1^{(1)}$ corresponding to polar ($c_1>0$) 
and weak ferromagnetic  ($0>c_1\ge c_1^{(1)}$) domains; at higher energies there could be 
other states. As $c_ 1$ is decreased further deep into ferromagnetic ($c_1<c_1^{(1)}$) domain, 
the axisymmetric vortex-bright  solitons are no longer the lowest-energy states. 
For $  c_1^{(1)} >c_1> c_1^{(2)}$, a new type of asymmetric soliton emerges  
with an energy lower than the axisymmetric soliton(s), which become excited 
states. Eventually, all types of states collapse for  $c_1< c_1^{(2)}$ 
because of an excess of attraction.  The numerical values of $c_1$, e.g. 
$c_1^{(1)}$ and   $c_2^{(1)}$, for the appearance of an asymmetric soliton  
for $c_1\le c_1^{(1)}$, and finally, its collapse for $c_1\le c_1^{(2)}$ 
depend on $c_ 0$ and $\gamma$. Using the phase-winding numbers (angular momentum) 
of the three-component wavefunction to denote a vortex \cite{Mizushima}, the 
axisymmetric vortex-bright  solitons  are classified as  $ (-1,0,+ 1)$ 
and $(0,+ 1,+ 2) \equiv (- 2,- 1, 0)$  solitons, where the numbers 
in the parenthesis are the phase-winding numbers of $\psi_{+1}$, $\psi_0$ 
and $\psi_{-1}$, respectively. Here the $\pm$ signs in the winding number denote
a vortex and and an anti-vortex rotating in opposite directions, respectively.
For example, the soliton $(-1,0,+1)$ denotes a state of angular momentum 
$\mp 1$ in components $\psi_{\pm 1}$ and angular momentum 0 in component 
$\psi_0$. Here, the cores of the vortices in $m_f= \pm 1$ components are 
occupied by the polar ($m_f = 0$) component, and thus these solitons can be 
termed  {\em polar-core} vortex-bright solitons. There are no stable stationary 
axisymmetric solitons of type $(0,0,0)$ without any angular momentum in all components. The details of a 
$(- 1,0,+ 1)$ vortex-bright  soliton  $-$ energy and density $-$ are independent 
of the value of $c_1$ $-$ positive or negative.  This is due to 
the fact that for the stable minimum energy solitons of this type spin density vector 
$\bf F$ is uniformly zero. However, the same of a $(0,+ 1,+ 2)$ vortex-bright  soliton 
and an asymmetric vortex-bright soliton  are dependent on  $c_1$. 
We will use a variational method to analytically study the axisymmetric  vortex-bright  solitons below. 

  Our numerical studies   show that the longitudinal magnetization 
${\cal M} = \int \{\rho_{+1}(\bf r)-\rho_{-1}(\bf r)\}d\bf r$ is zero for the  $(- 1,0,+ 1)$ solitons; 
whereas it can be non-zero for  the $(0,+ 1,+ 2)$ solitons. 
This guides our choices of simple variational {\em ansatz}
to model the vortex-bright  solitons.
 The $(-1,0,+1)$ vortex-bright soliton  with zero magnetization $\cal M$ can be analyzed 
using the following variational {\em ansatz}
\begin{align}
\psi_{\pm 1} &= \frac{A_1 r}{\sigma_1^2} \exp\left( - \frac{r^2}{2 \sigma_1^2}\mp i \phi\right),\label{ansatz1}\\
\psi_0&=i \frac{A_2}{\sigma_2} \exp\left(-\frac{r^2}{2 \sigma_2^2}\right)\label{ansatz2},
\end{align}
where $r = \sqrt{x^2+y^2}$ and $\phi = \tan^{-1}(y/x)$ are the radial and azimuthal
coordinates, $A_i,\sigma_i$ are the variational parameters, which denote the amplitude 
and the width of the component wavefunctions, respectively. The condition of zero magnetization fixes the amplitudes of
components { $\psi_{\pm 1}$} to be equal. The equal and opposite phases $(\mp \phi)$ of these components 
guarantee their opposite directions of rotation with unit angular momentum $-$ vortex and anti-vortex.
 Only three of the variational parameters are independent of each other as the fourth, 
say $A_2$, is fixed by the normalization ($=1$). The  variational energy of the soliton,
 obtained by substituting Eqs. (\ref{ansatz1}) and (\ref{ansatz2}) in 
Eq. (\ref{energy}), is
\begin{eqnarray}
E &=& \frac{\pi}{2} \biggr[\biggr\{\frac{A_2^2}{\sigma_2^2}+\frac{4 A_1^2} {\sigma_1^2}-\frac{16 \sqrt{2} A_1 A_2 
\gamma  \sigma_1^2 \sigma_2}
{\left(\sigma_1^2+\sigma_2^2\right)^2}\biggr\}\nonumber \\
&+&c_0 \biggr\{\frac{A_1^4 }  {\sigma_1^2}+\frac{4 A_1^2 A_2^2  \sigma_2^2}{\left(\sigma_1^2
+\sigma_2^2\right)^2}+\frac{A_2^4 }{2 \sigma_2^2} \biggr\} \biggr],
\label{E1}
\end{eqnarray}
where $A_2$ is determined by the normalization constraint: 
\begin{equation}
A_2 = \frac{\sqrt{1-2 \pi A_1^2}}{\sqrt{\pi } }\label{const}.
\end{equation}
As mentioned earlier, in this case  $|{\bf F}|^2=0$;  consequently, variational  energy (\ref{E1}) is independent of $c_1$.
Energy  (\ref{E1}) can be minimized with respect to the variational parameters 
$A_i$ and  $\sigma_i$, with Eq. (\ref{const}) acting as a constraint, 
to determine $A_i$ and $\sigma_i$.
The numerical result for the component wave functions of a stationary $(-1,0,+1)$ vortex-bright  soliton 
is obtained by an imaginary-time 
simulation of Eqs. (\ref{gps-1}) and (\ref{gps-2}) with an initial guess of component 
wave functions (\ref{ansatz1}) and (\ref{ansatz2}). { In case of axisymmetric
$(-1,0,+1)$ ground state solutions, ansatz  (\ref{ansatz1}) and 
(\ref{ansatz2}) ensure faster convergence of the numerical results as compared to Gaussian initial 
guess for the three-component wavefunction, which also lead to the same final converged solutions.} 

Next we consider the axisymmetric  $(0,+1,+2)$ vortex-bright soliton, which has higher energy 
than an axisymmetric  $(- 1,0,+ 1)$  vortex-bright soliton with the same parameters 
$c_0,c_1,$ and $\gamma$. For a variational study of the $(0,+1,+2)$ vortex-bright soliton,
we adopt the following variational {\em ansatz}
\begin{align}\label{v1}
\psi_{+1}&=i \frac{A_1}{\sigma_1} \exp\left(-\frac{r^2}{2 \sigma_ 1^2}\right)\\
\psi_0&=\frac{A_2}{\sigma_2^2} r\exp\left(-\frac{r^2}{2 \sigma_2^2}+i \phi \right)\\
\psi_{-1}&=-i\frac{A_3}{\sigma_3^3} r^2\exp\left(-\frac{r^2}{2 \sigma_3^2}+i2 \phi \right), \label{v3}
\end{align}
where $A_i$ and $\sigma_i$ are the variational parameters  for the amplitude and the width of the component wave functions.  The phases $0,\phi$, and $2\phi$ of the components $\psi_{+1}, \psi_0,$ and
$\psi_{-1}$, respectively, ensure their angular momenta as $(0,+1,+2)$,
In the case of a $(- 1,0, +1)$ vortex-bright soliton, the  zero magnetization condition 
(${\cal M}=0$) fixes the amplitudes of the 
wave function components $\psi_{\pm 1}$ to be equal. {Our numerical simulations confirm that} this is not the case for a
stable  $(0,+1,+2)$ vortex-bright soliton where, in general, ${\cal M} \ne 0$. 
{ Hence, a fixed norm (=1) is the only constraint, which 
reduces the number of independent variational parameters (=6)  by one.}  
Using Eqs. (\ref{v1})-(\ref{v3}), the energy (\ref{energy}) of the soliton can be written as
\begin{widetext}
\begin{align}
E &= \frac{1}{8} \pi  \left[4 \left\{\frac{A_1^2}{\sigma_1^2}+2 \frac{A_2^2}{ \sigma_2^2}
-\frac{8 \sqrt{2} A_1 A_2 \gamma \sigma_1 \sigma_2^2}
{\left(\sigma_1^2+\sigma_2^2\right)^2}+6 \frac{A_3^2} {\sigma_3^2}-\frac{32 \sqrt{2} A_2A_3 \gamma  \sigma_2^2 \sigma_3^3}
{\left(\sigma_2^2+\sigma_3^2\right)^3}\right\}\right.\nonumber\\
&+c_0\left\{2 \frac{A_1^4} {\sigma_1^2}+3 \frac{A_3^4} {\sigma_3^2}+\frac{48 A_2^2 A_3^2 \sigma_2^4 \sigma_3^2}
{\left(\sigma_2^2+\sigma_3^2\right)^4}+A_1^2 \left(\frac{8 A_2^2 \sigma_1^2}{\left(\sigma_1^2+\sigma_2^2\right)^2}
+\frac{16 A_3^2 \sigma_1^4 }{\left(\sigma_1^2+\sigma_3^2\right)^3}\right)+\frac{A_2^4} {\sigma_2^2}\right\} \nonumber\\
&+c_1\left.\left\{2 \frac{A_1^4} {\sigma_1^2}
+3 \frac{A_3^4} {\sigma_3^2}
+
\frac{48 A_2^2 A_3^2 \sigma_2^4 \sigma_3^2}{\left({\sigma_2^2}
+{\sigma_3^2}\right)^4}
+A_1^2 
\left(\frac{8 A_2^2\sigma_1^2}{\left({\sigma_1^2}+{\sigma_2^2}\right)^2}-
\frac{16 A_3^2 \sigma_1^4}{\left({\sigma_1^2}+{\sigma_3^2}\right)^3}\right)
+\frac{256 A_1 A_2^2 A_3\sigma_1^5 \sigma_2^2 \sigma_3^3}{\left(
\sigma_2^2 \sigma_ 3^2+ 2 \sigma_1^2 \sigma_ 3^2+ \sigma_1^2 \sigma_ 2^2
\right)^3}\right\} \right]\label{E2}.
\end{align}
\end{widetext}
The condition of 
fixed norm $(=1)$ leads to one constraint relating the variational 
parameters 
$A_i$ and $\sigma_i$:
\begin{eqnarray}
 \pi (A_1^2 +A_2^2 +2 A_3^2) &=& 1 \label{const2}.
\end{eqnarray}
Minimizing energy  (\ref{E2}) with respect to the variational parameters $A_i$ and $\sigma_i$ under the constraint
(\ref{const2}), one can determine the variational parameters. 

For a $(-2,-1,0)$ vortex-bright soliton {which is degenerate with a (0,+1,+2) soliton}, 
the appropriate variational ansatz can be obtained from Eqs. (\ref{v1})-(\ref{v3}) by transformations 
$\psi_{m_f}(r,\phi) \rightarrow \psi_{-m_f}(r,-\phi)$ or $\psi_{m_f}(x,y) \rightarrow \psi_{-m_f}(x,-y)$. The
degeneracy of these states is due to underlying symmetry of Eqs. (\ref{gps-1})-(\ref{gps-2}), which remain
invariant under the transformation $y \rightarrow -y$ and $\psi_{m_f}(x,y)\rightarrow \psi_{-m_f}(x,-y)$. 
Under this transformation, a $(-1,0,+1)$ vortex-bright soliton transforms into itself as can be
confirmed from the variational ansatz (\ref{ansatz1})-(\ref{ansatz2}); hence there is no degenerate counterpart
for a $(-1,0,+1)$ vortex-bright soliton.  Equations  (\ref{v1})-(\ref{v3})
are invariant under simultaneous transformations of $\gamma\rightarrow -\gamma$ and 
$\psi_0\rightarrow\psi_0 e^{i\pi}$ (while keeping $\psi_{\pm 1}$ unchanged). It implies
that for negative  $\gamma$, the vortex-bright solitons are fundamentally
identical to those for   positive   $\gamma$ except for a phase difference 
of $\pi$ in their $m_f =0$ components.

\section{Numerical Procedure}
\label{Sec-III}
  
The coupled equations  (\ref{gps-1})-(\ref{gps-2}) can be solved 
by time-splitting 
Fourier Pseudo-spectral method \cite{psanand}  and time-splitting Crank-Nicolson method  \cite{Wang, Bao, Muruganandam}. Here, we  extend the Fourier Pseudo-spectral method 
to the coupled GP equations with SO coupling terms and use the same to solve Eqs.  (\ref{gps-1})-(\ref{gps-2}). 
The coupled set of GP equations (\ref{gps-1})-(\ref{gps-2}) 
can be represented in a simplified form as
\begin{equation}
\frac{i\partial \Psi}{\partial t} = \left({H_1+H_2+H_3}\right)\Psi\label{GPE},
\end{equation}
where $\Psi=(\psi_{+1},\psi_{0},\psi_{-1})^T$ {with $T$ denoting the transpose}, $H_1$, $H_2$ and $H_3$
are $3\times3$ matrix operators defined as
\begin{align}
H_1 &= 
\begin{pmatrix} 
 {\cal H}+c_1(\rho_0+\rho_{-}) & 0 &0 \\
0 & {\cal H}+c_1\rho_{+}&0 \\
0 & 0 & {\cal H}+c_1(\rho_0-\rho_{-})
\end{pmatrix},\\ 
H_2 &= 
\begin{pmatrix} 
 0 & c_1\psi_0\psi_{-1}^*&0 \\
c_1\psi_0^*\psi_{-1} & 0 &\psi_0^*\psi_{+1} \\
0 & c_1\psi_0\psi_{+1}^* & 0
\end{pmatrix}, \\
H_3 &=-i\frac{\gamma}{\sqrt{2}} \begin{pmatrix} 
 0 & \partial_-&0 \\
\partial_+ & 0 & \partial_- \\
0 & \partial_+   & 0
\end{pmatrix},
\end{align}
where
\begin{align}
\rho_{\pm}&=\rho_{+1}\pm \rho_{-1}, \quad  \partial_{\pm}= \left(\frac{\partial}{\partial x} \pm i \frac{\partial}{\partial y}\right)
\end{align}

Now, the lowest order time-splitting involves solving the 
following equations successively
\begin{eqnarray}
\frac{i\partial\Psi}{\partial t} &=& H_1\Psi,\label{GPE1}\\
\frac{i\partial\Psi}{\partial t} &=& H_2\Psi,\label{GPE2}\\
\frac{i\partial\Psi}{\partial t} &=& H_3\Psi\label{GPE3}.
\end{eqnarray}
Eq. (\ref{GPE1}) can be numerically solved using Fourier Pseudo-spectral method 
\cite{Wang} which we employ in this paper or semi-implicit Crank-Nicolson method \cite{Muruganandam} {and involves
additional time-splitting of $H_1$ into its spatial derivative and non-derivative parts.} 
The numerical solutions of Eq. (\ref{GPE2}) have been discussed in Refs. \cite{Wang,Martikainen}.  
We use Fourier Pseudo-spectral method to accurately solve Eq. (\ref{GPE3}).
 In Fourier space, Eq. (\ref{GPE3}) is
\begin{equation}
\frac{i\partial\tilde{\Psi}}{\partial t} = \tilde{H}_3\tilde{\Psi}\label{GPEF3},
\end{equation}
where tilde indicates that the quantity has been Fourier transformed.
 Hamiltonian $H_3$ in Fourier space is
\begin{equation}
\tilde{H}_3 = -i\frac{\gamma}{\sqrt{2}} \begin{pmatrix} 
 0 & ik_x+k_y& 0\\
 ik_x-k_y  &0 & ik_x+k_y \\
0 &  ik_x-k_y & 0  
\end{pmatrix}
\end{equation}
The solution of Eq. (\ref{GPEF3}) is
\begin{align}
\tilde{\Psi}(t+dt) &= e^{-i\tilde{H}_3 dt} \tilde{\Psi}(t) = e^{-i\hat{O}} \tilde{\Psi}(t),\\
                   &= \left(I + \frac{\cos{\Omega}-1}{\Omega^2}\hat{O}^2-
i\frac{\sin{\Omega}}{\Omega}\hat{O}\right)\tilde{\Psi}(t),\label{GPEF3S}
\end{align}
where $\Omega = \sqrt{|A|^2 +|B|^2}$, where $A = -i\frac{\gamma}{\sqrt{2}}\left(ik_x + k_y\right)dt$ 
and $ B = -i\frac{\gamma}{\sqrt{2}}\left(ik_x -k_y\right)dt$, and $\hat{O}$ is
defined as
\begin{equation}
\hat{O} = \begin{pmatrix}
0 & A & 0\\
A^*&0& B^*\\
0&B&0
\end{pmatrix}.
\end{equation}
The wavefunction in Eq. (\ref{GPEF3S}) is in Fourier space and can be inverse Fourier
transformed to obtain the solution  in configuration  space.
In this study, in  space and time discretizations, we use space and time steps of $0.1$ and $0.005$, 
respectively, in imaginary-time simulation, whereas in real-time simulation
 these are, respectively, $0.1$ and $0.0005$.

\section{Numerical results}
\label{Sec-IV}

\begin{figure}[t]

\begin{center}

\includegraphics[trim = 0mm 0mm 0mm 0mm, clip,width=0.475\linewidth,clip]{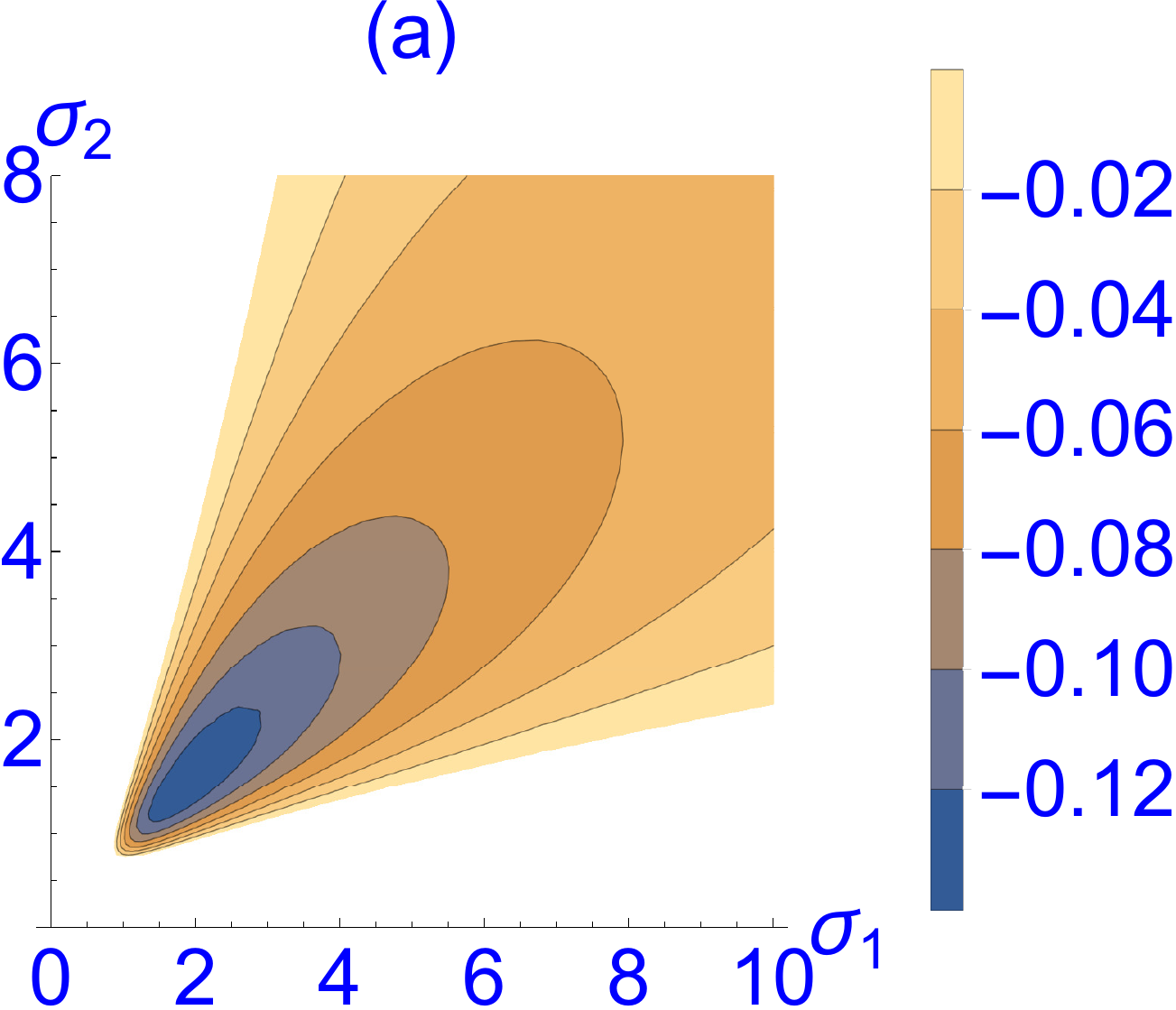}
\includegraphics[trim = 0mm 0mm 0mm 0mm, clip,width=0.51\linewidth,clip]{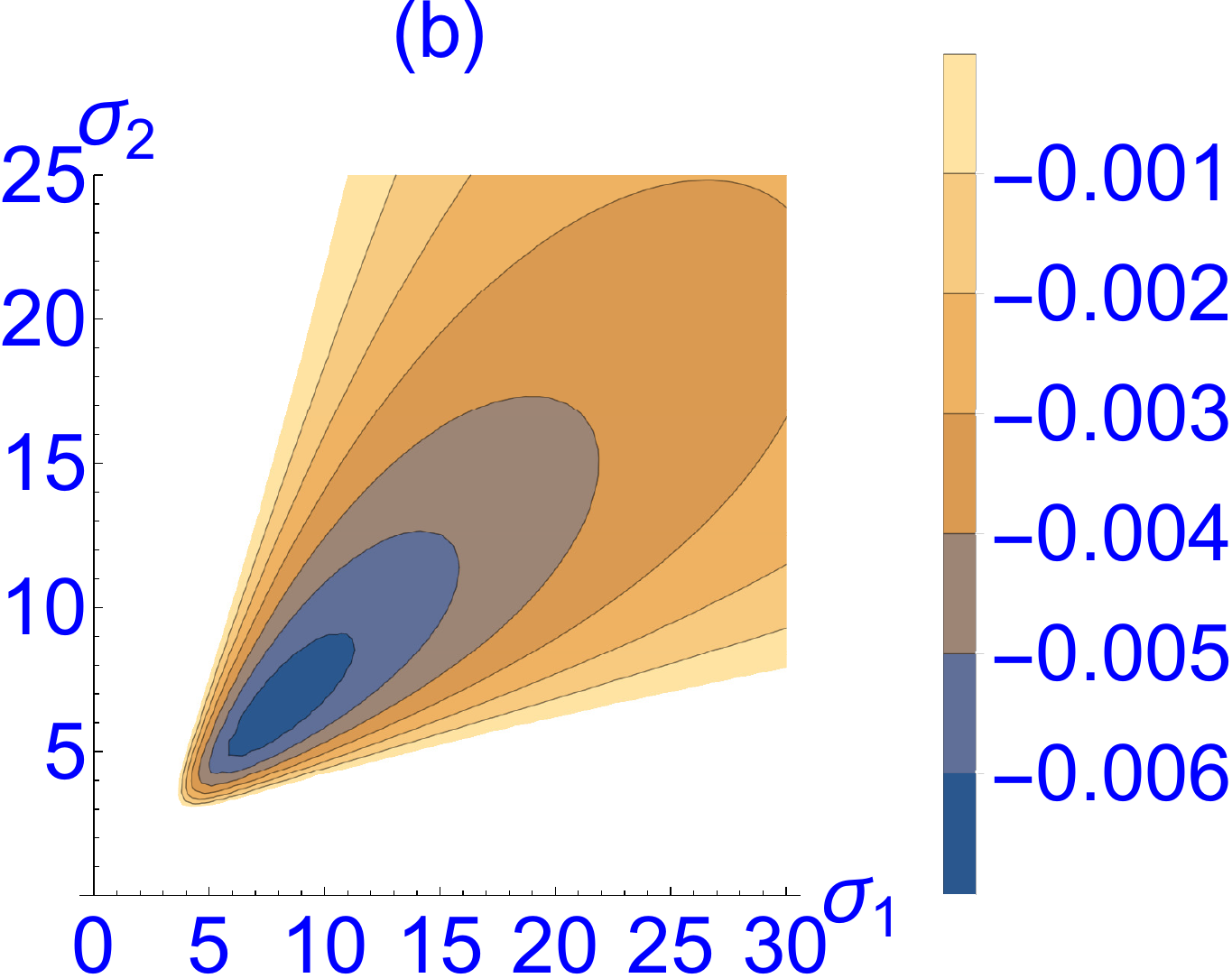}
 
\caption{ (Color online) Two-dimensional contour plot of energy $E$ of Eq. (\ref{E1}) as a function of widths $\sigma_1$ and $\sigma_2$   for (a) $c_0=-4, \gamma=0.5$ and (b) $c_0=-5, \gamma=0.1$. The actual values of $A_1$ and $A_2$ corresponding to the energy minima in these two cases 
have been used in Eq. (\ref{E1}) to prepare these plots. }
\label{fig0}

\end{center}

\end{figure}

How the SO coupling creates a stable 2D soliton is explicit in the expression for energy (\ref{E1}). 
A stable bound soliton corresponds to a global minimum of the variational energy. In fact, for SO coupling $\gamma=0$,  this energy expression is positive and tends to zero as $\sigma_1,\sigma_2 \to \infty$ and does not have 
any minimum for $c_0\ge -7$, beyond which $E\to -\infty$ as $\sigma_1,\sigma_2\to 0$ and the system collapses.  
The contribution of the SO coupling to energy $E$ of Eq. (\ref{E1}) is always negative in the form of a shallow well in the $\sigma_1 -\sigma_2$
plane.  Hence by choosing $c_0>-7$ (collapse-free region) and an adequate value of SO coupling $\gamma$, one can have a global minimum at negative 
energy in the energy expression (\ref{E1}) as a function of $\sigma_1$ and $\sigma_2$ corresponding to a stable 2D soliton.   
To illustrate how the SO coupling leads to a energy minimum we consider two examples: (a) $c_0=-4, \gamma=-0.5$ and (b) 
 $c_0=-5, \gamma=-0.1$. In these two cases there is a energy minimum (a) $E_{\mathrm{min}}=-0.1441, \sigma_1=2.00, \sigma_2
=1.588, A_1=0.2424$ and (b) 
 $E_{\mathrm{min}}=-0.00668, \sigma_1=8.177, \sigma_2
=6.587, A_1=0.221$. How the energy varies as a function of widths $\sigma_1$  and $\sigma_2$ 
for  given amplitudes $A_1$ and $A_2$ can be seen in  contour plots of energy as a function of
the widths with the given amplitudes. These plots are shown in Figs. \ref{fig0}(a) and (b) in these 
two cases explicitly showing the global minima of energy at negative energies. Outside the shaded areas in these plots, the 
energy function is zero or positive. The same thing also happens in the 
three-component energy function (\ref{E2}), which, however, is difficult to illustrate graphically.

\subsection{Axisymmetric vortex-bright soliton}
\label{Sec-IVA} 

\begin{figure}[t]
\begin{center}
\includegraphics[trim = 5mm 0mm 4mm 0mm, clip,width=0.49\linewidth,clip]{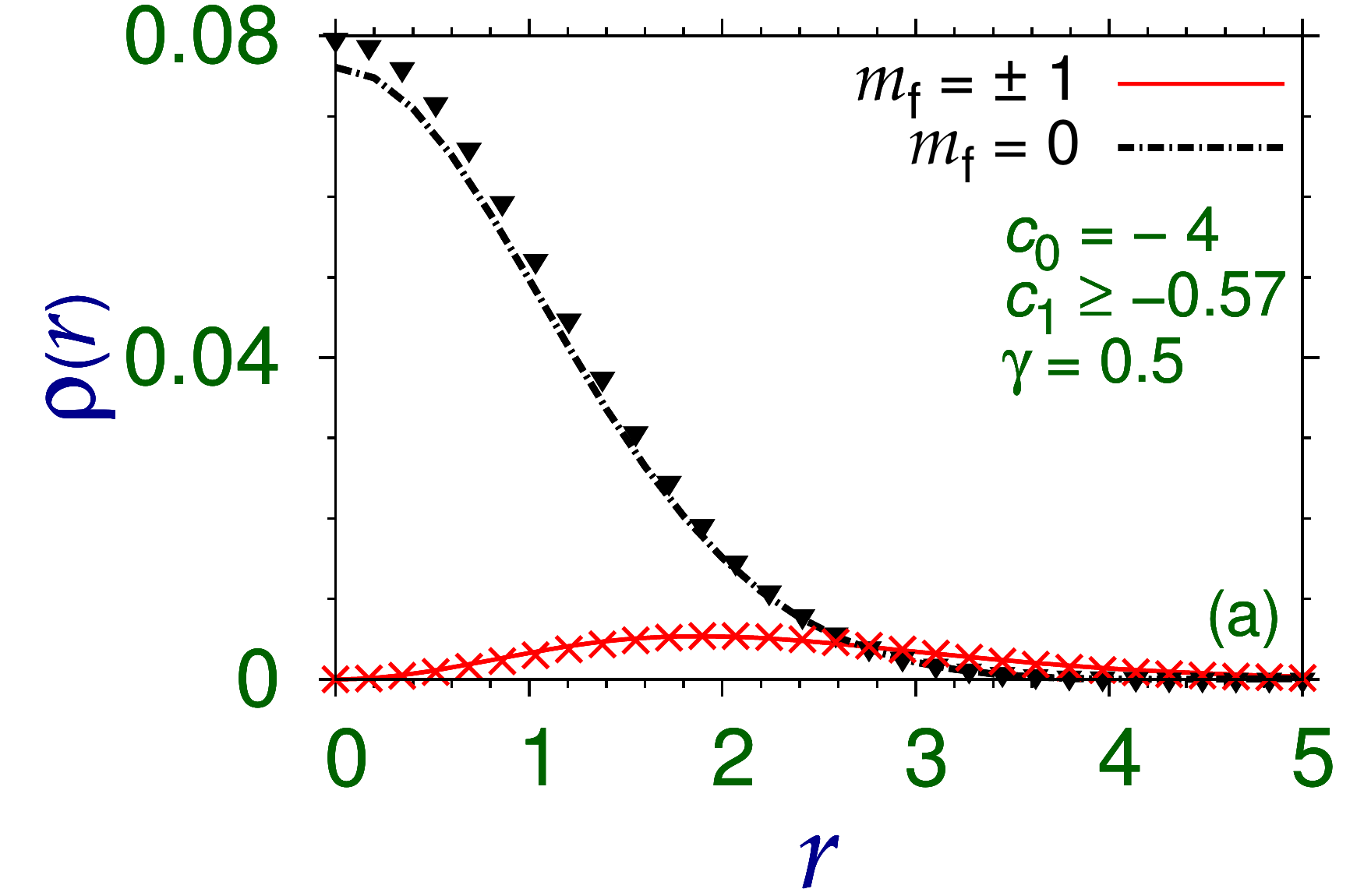}
\includegraphics[trim = 5mm 0mm 4mm 0mm, clip,width=0.49\linewidth,clip]{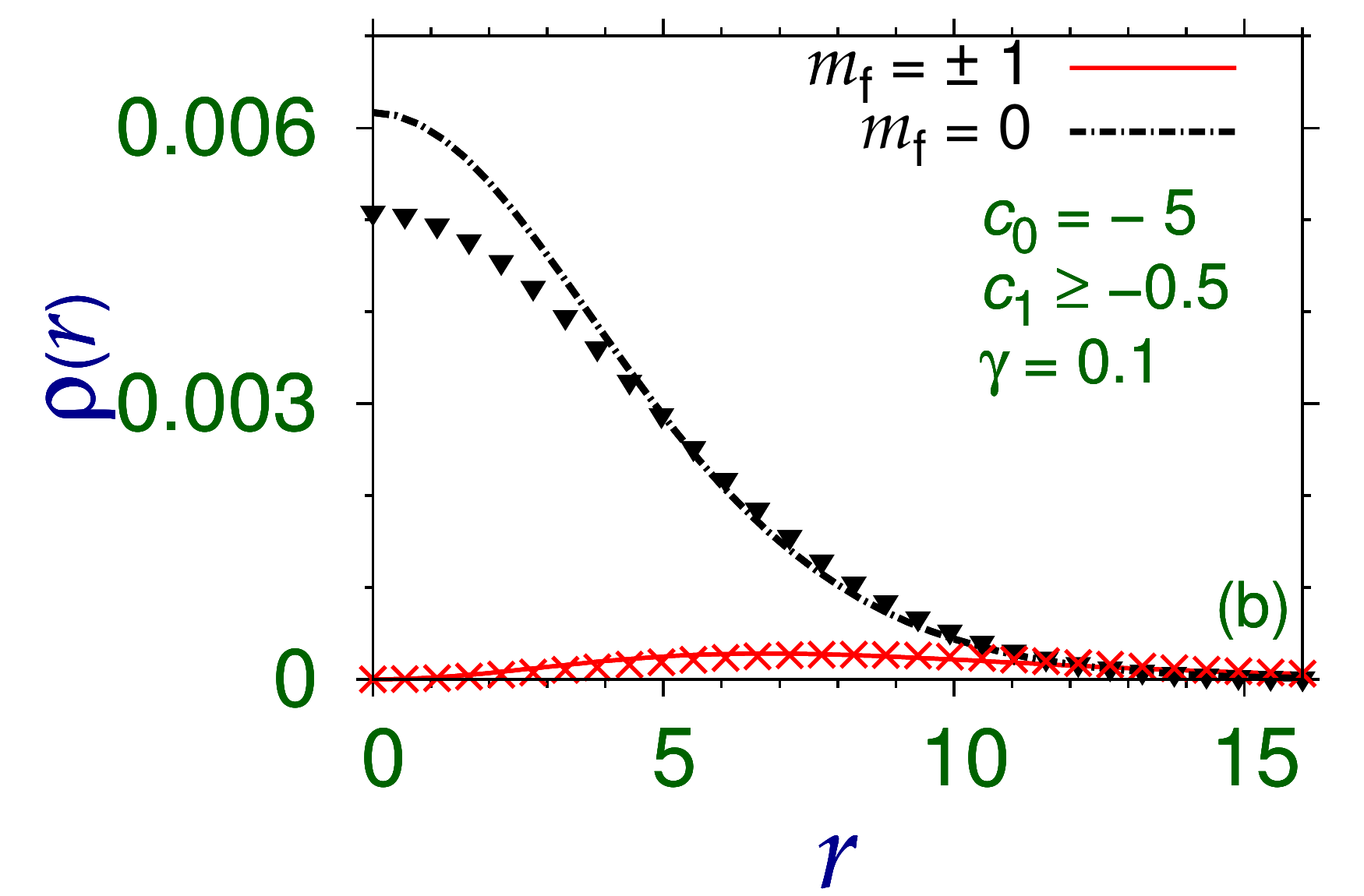}
\includegraphics[trim = 5mm 0mm 4mm 0mm, clip,width=0.49\linewidth,clip]{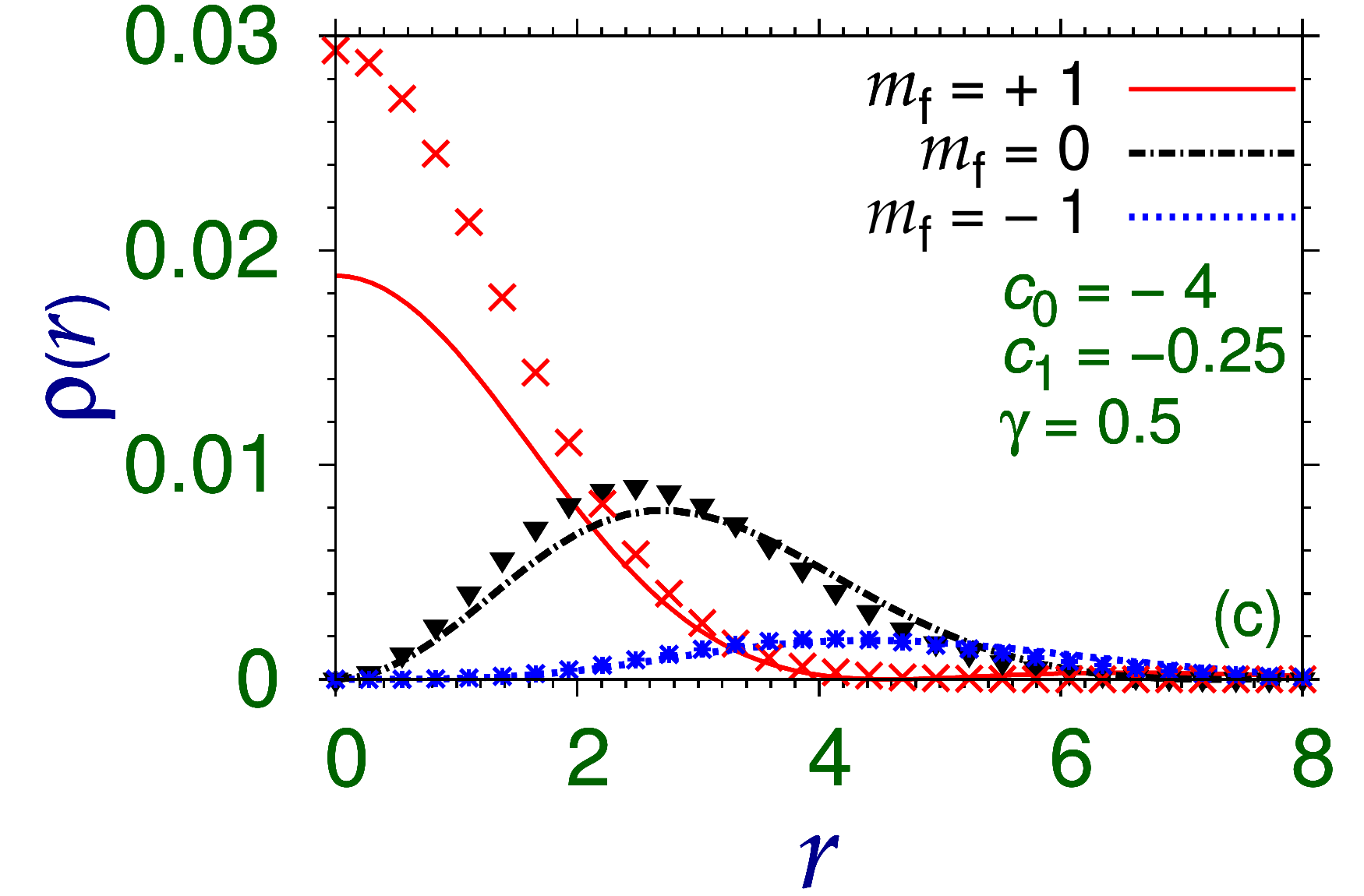}
\includegraphics[trim = 5mm 0mm 4mm 0mm, clip,width=0.49\linewidth,clip]{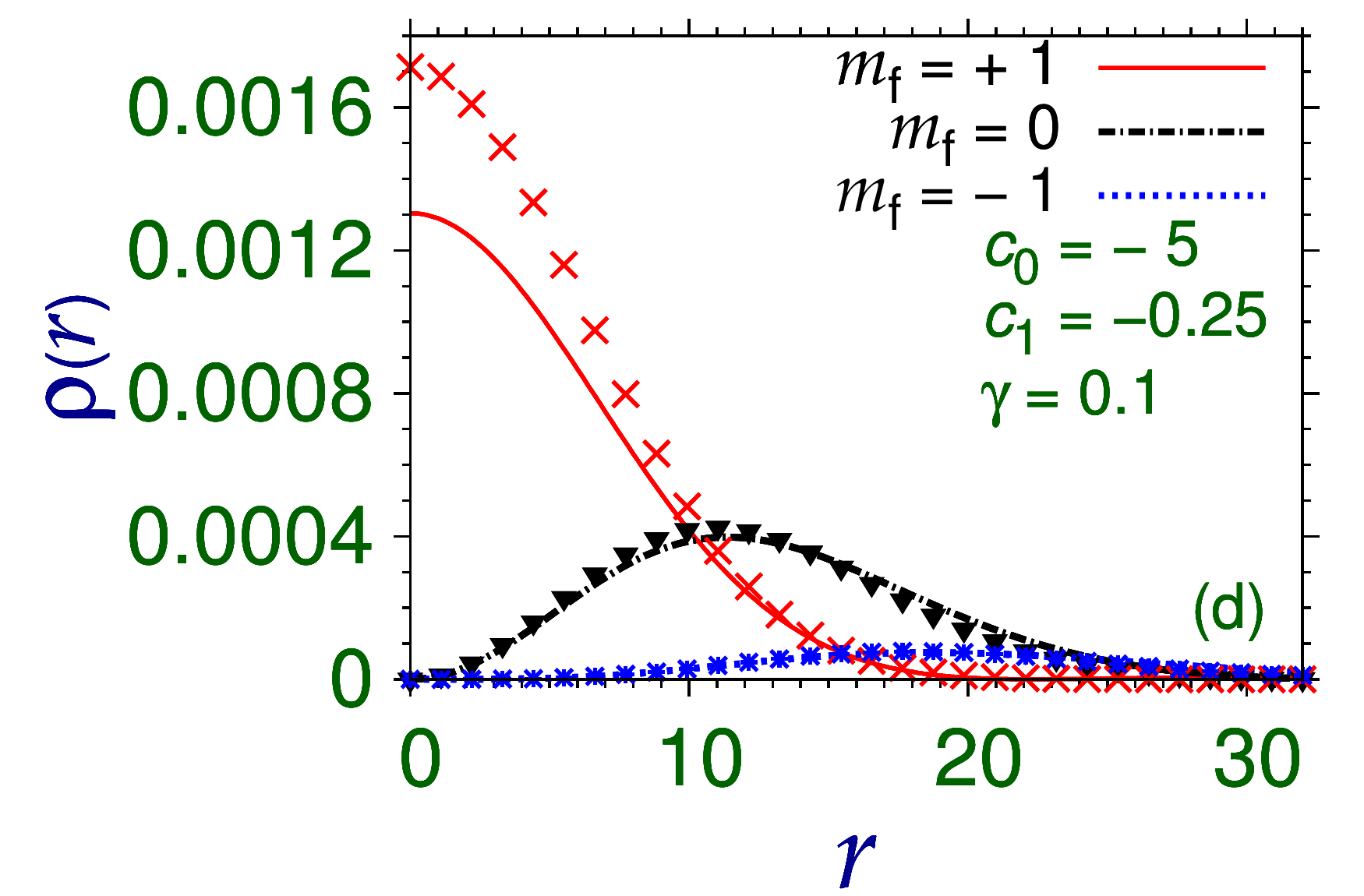}
\caption{(Color online) Numerical (line) and variational 
(chain of symbols) results for radial density $\rho(r)$ versus $r$  of the components 
in an axisymmetric $(- 1,0,+ 1)$ vortex-bright soliton  for (a)  $c_0 = -4, {c_1\ge -0.57}$, and  
$\gamma = 0.5$ and for (b) $c_0 = -5, {c_1\ge -0.5}$, and  $\gamma = 0.5$. The same in an axisymmetric 
$(0,+ 1,+ 2)$ vortex-bright soliton  for (c)  $c_0 = -4, c_1=-0.25$, and  $\gamma = 0.5$ and for 
(d) $c_0 = -5, c_1=-0.25$, and  $\gamma = 0.5$. All quantities in this and following figures are dimensionless.}
\label{fig2}
\end{center}
\end{figure}

 The numerical and analytic variational  results  for radial density $\rho(r)$ versus $r$ 
for an axisymmetric $(- 1,0,+ 1)$ vortex-bright soliton  for (a) $c_0 = -4, c_1 \ge -0.57$, 
and $\gamma = 0.5$ and for (b) $c_0 = -5, {c_1 \ge -0.5 }$, and {$\gamma = 0.1$} are
shown in Figs. \ref{fig2}(a) and (b), respectively. The numerical result is obtained by an imaginary-time 
simulation of Eqs. (\ref{gps-1}) and (\ref{gps-2}) with the initial guess of component 
wave functions (\ref{ansatz1})-(\ref{ansatz2}). 
The numerical and analytic variational  results  for radial density   $\rho(r)$ versus $r$ 
for the axisymmetric  $(0,+ 1,+ 2)$ vortex-bright solitons for the same {$c_0$, $\gamma$ 
and $c_1 = -0.25$}
are   shown in Figs. \ref{fig2}(c) and (d).
The numerical result in this case is obtained by an 
imaginary-time 
simulation of Eqs. (\ref{gps-1}) and (\ref{gps-2}) with the initial guess of component 
wave functions (\ref{v1})-(\ref{v3}). 
The wave function components $\psi_{+1}, \psi_0,$ and $\psi_{-1}$ in Figs. \ref{fig2}(a) and (b)
carry angular momenta  $- 1, 0$ and $+ 1$  respectively, whereas in Figs. \ref{fig2}(c) 
and (d), they  carry angular momenta $0,+ 1$ and $+ 2$.
The $(- 1,0,+ 1)$ states of Fig. \ref{fig2}(a) and (b) 
are the ground states of the system, whereas the $(0,+1,+2)$  states of Fig. \ref{fig2}(c) and (d) are  excited states.  
For $ c_0=-4, \gamma =0.5$, viz. Figs. \ref{fig2}(a) and (c), the axisymmetric $(- 1,0,+ 1)$
vortex-bright soliton is the ground state for $c_1 \ge c_1^{(1)}= -0.57$ and for 
$ c_0=-5, {\gamma =0.1}$, viz. Figs. \ref{fig2}(b) and (d), the axisymmetric $(- 1,0,+ 1)$ vortex-bright soliton 
is the ground state for $c_1 \ge c_1^{(1)}= -0.5$.

\begin{figure}[t]
\begin{center}
\includegraphics[trim = 10mm 0mm 10mm 0mm, clip,width=\linewidth,clip]{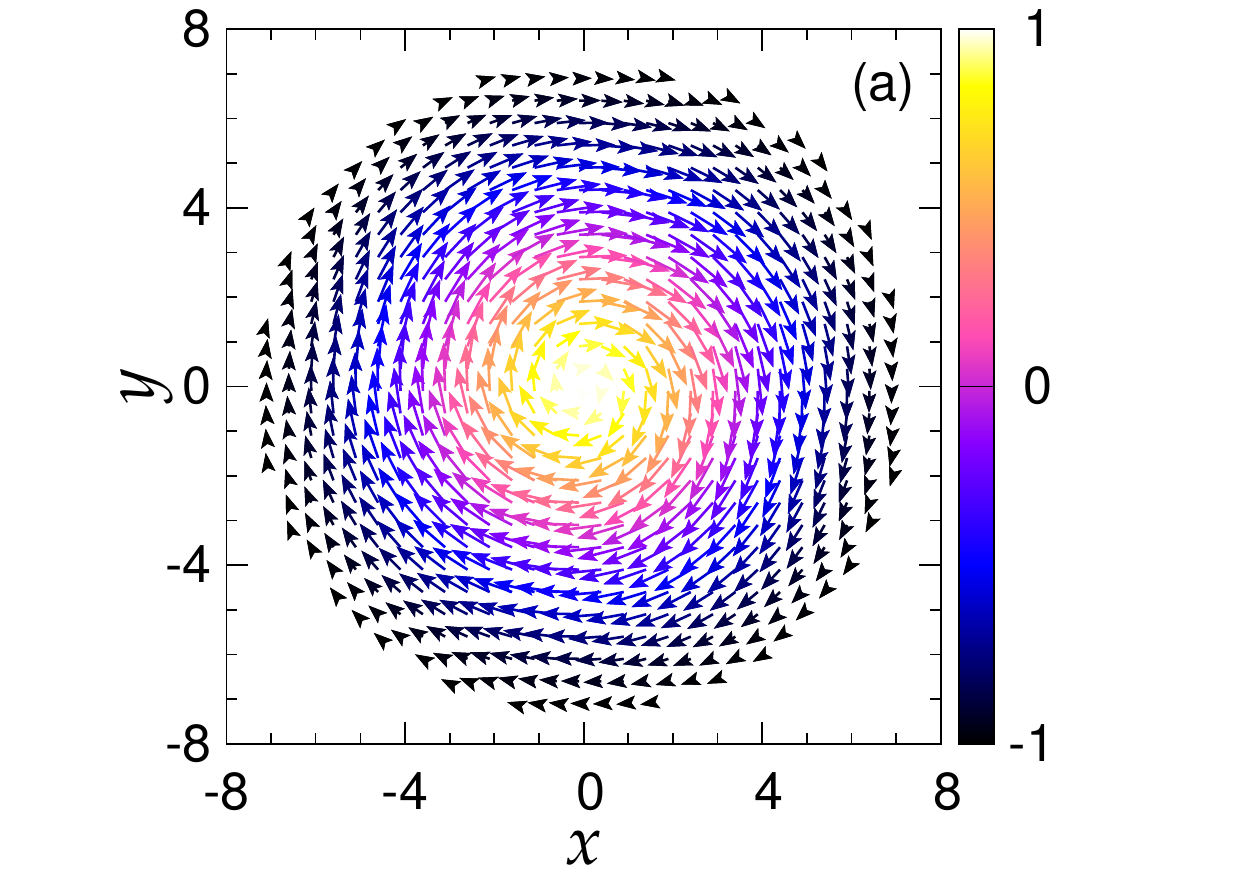} 
\includegraphics[trim = 4mm 0mm 0mm 0mm, clip,width=0.49\linewidth,clip]{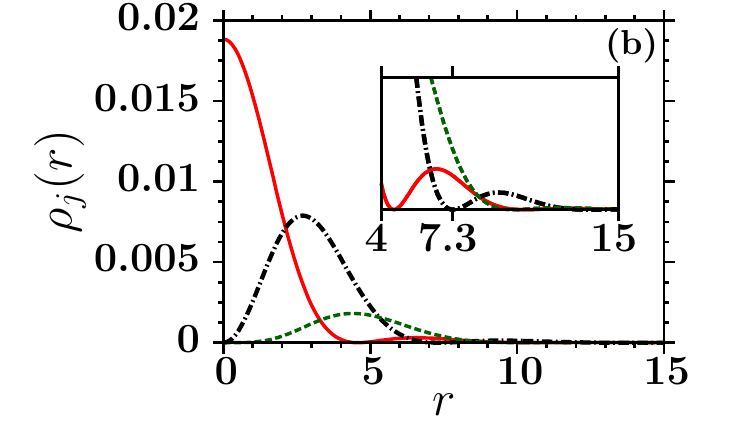} 
\includegraphics[trim = 4mm 0mm 0mm 0mm, clip,width=0.49\linewidth,clip]{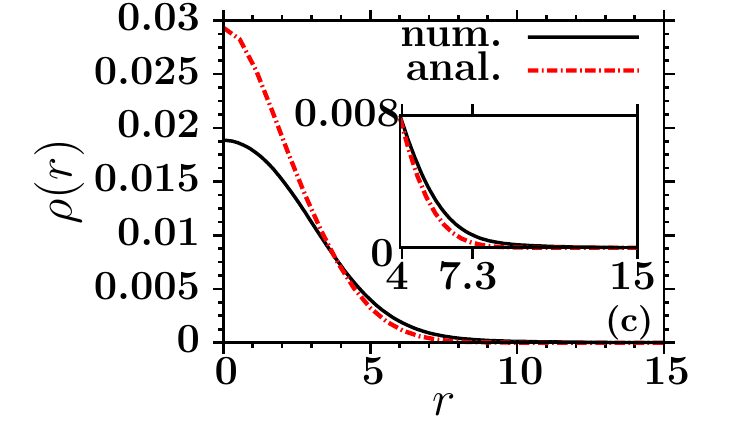} 
\caption{ (Color online) (a)  The projection of local magnetization vector (normalized to unity) 
on the $x-y$ plane for the axisymmetric $(0,+1,+2)$ vortex-bright  soliton of 
Fig. \ref{fig2}(c) with $c_0=-4, c_1=-0.25, \gamma = 0.5$. The color indicates the $l_z$ component.
At the center, color value of $+1$ indicates that the local magnetization vector is
directed along $+z$ axis, similarly at the edge, color value of $-1$ indicated that
the local magnetization vector is directed along $-z$ axis. {(b) Numerical results
for component densities $\rho_j(r)$ versus $r$ for the axisymmetric $(0,+1,+2)$ vortex-bright  soliton
with $c_0=-4, c_1=-0.25, \gamma = 0.5$ showing the oscillation in density; here solid red,
dot-dashed black and dashed green lines show the densities of $m_f = +1$, $m_f = 0$ and $m_f = -1$
components, respectively and the inset shows the zoom-in of main figure from $r =4$ to $r=15$. (c) Total numerical (num.)
and variational or analytic (anal.) densities  as a function of $r$ for the axisymmetric $(0,+1,+2)$ 
vortex-bright  soliton of Fig. \ref{fig2}(c); inset shows the same above the cross-over between
the numerical and variational curves.
}
}
\label{fig-3}
\end{center}
\end{figure}

To describe the spatial orientation of the local magnetization vector in a 
spinor vortex BEC it is convenient to define  a local magnetization vector 
$\mathbf l$, which points in the direction of spin,  as the cross product of two 
vectors $\mathbf m$ and $\mathbf n$ \cite{Mizushima}
\begin{equation}
\bf l = \bf m \times \bf n,
\end{equation}
where ${\bf m} \equiv (m_x,m_y,m_z) = {\rm Re}(\psi_x,\psi_y,\psi_z)$ and 
${\bf n} \equiv  (n_x,n_y,n_z) = {\rm Im}(\psi_x,\psi_y,\psi_z)$,
and
\begin{eqnarray}
\psi_x &=& \frac{-\psi_{+1} +\psi_{-1}}{\sqrt{2}},\\
\psi_y &=& \frac{-i(\psi_{+1} +\psi_{-1})}{\sqrt{2}},\\
\psi_z &=& \psi_0,
\end{eqnarray}
where ${\rm Re}$ and ${\rm Im}$ stand for real and imaginary parts, respectively,
$\psi_x, \psi_y,\psi_z$ are the components of the order parameter in Cartesian basis \cite{Kawaguchi}.
 An axisymmetric  $(0,+1,+2)$ vortex can have two
distinct spin textures   \cite{Mizushima} which are  the spatial distribution of the local magnetization
vector.
For the $(0,+1,+2)$ vortex, the unit vector $\hat {\bf l} = \hat z \cos \beta(r)+\sin \beta(r)(\hat x\cos\phi+\hat y \sin\phi ),$
here $\phi$ is the azimuthal angle, and $\beta(r)$ varies from $\beta(0) = 0 $ to
$\beta(R) = \pi/2$  for a Mermin-Ho coreless vortex \cite{mh} and  from 
$\beta(0) = 0 $ to
$\beta(R) = \pi$  for a
Anderson-Toulouse coreless vortex \cite{at}, where 
 the outer edge of the condensate is at $r=R$ \cite{Mizushima}. 
In Fig. \ref{fig-3}(a), we plot the numerically obtained projection of the local magnetization 
vector on the $xy$ plane for the axisymmetric  $(0,+1,+2)$ vortex-bright soliton shown in Fig, \ref{fig2}(c) 
for parameters $c_0=-4, c_1=-0.25, $ and $\gamma=0.5$. The color of the arrows in Fig. \ref{fig-3}(a), with value ranging from -1 to 1,
represents the $z$ component of the local magnetization vector. Here the spin texture
has been shown from origin to the second zero of $\rho_0$ which occurs
at $r=7.3$ in this case as in shown in Fig. \ref{fig-3}(b).  
It is evident from Fig. \ref{fig-3}(a) that the spin texture associated with axisymmetric (0,+1,+2)  
vortex-bright soliton is consistent with the spin texture of an Anderson-Toulouse  
coreless vortex \cite{Mizushima}. We find that from origin to the second zero of $\rho_0$
 a (0,+1,+2) vortex-bright
soliton   always has this spin texture.
At the origin (the first zero of $\rho_0$), the spin points along positive $z$ direction and
it gets fully inverted at the second zero of $\rho_0$. From the inset of Fig. \ref{fig-3}(b),
it is also evident that the densities of the three components actually have oscillations
with multiple zeros. This  oscillation and the consequent deviation from the Gaussian shape in density 
is the main reason for the difference in the analytic and variational density profiles shown in 
Fig. \ref{fig2}. Since the total norm  $(2\pi\int r\rho(r)dr=1)$ for both the numerical and variational densities is the same, 
it implies that the total variational density which is consistently larger than the numerical density
near the origin, as can be inferred from Figs. \ref{fig2}(a)-(d), must be smaller than the total numerical density
after a cross-over point. This is indeed the case for all the vortex-bright solitons shown
in Fig. \ref{fig2}. To illustrate it for the vortex-bright soliton shown in Fig. \ref{fig2}(c), 
the total numerical and variational densities are shown in Fig. \ref{fig-3}(c); the inset shows the densities 
in the domain where total numerical density is consistently higher than the variational one.

\subsection{Asymmetric solitons}
\label{Sec-IVB}

\begin{figure}[t]
\begin{center}
\includegraphics[trim = 0cm 0cm 0cm 0cm, clip,width=\linewidth,clip]{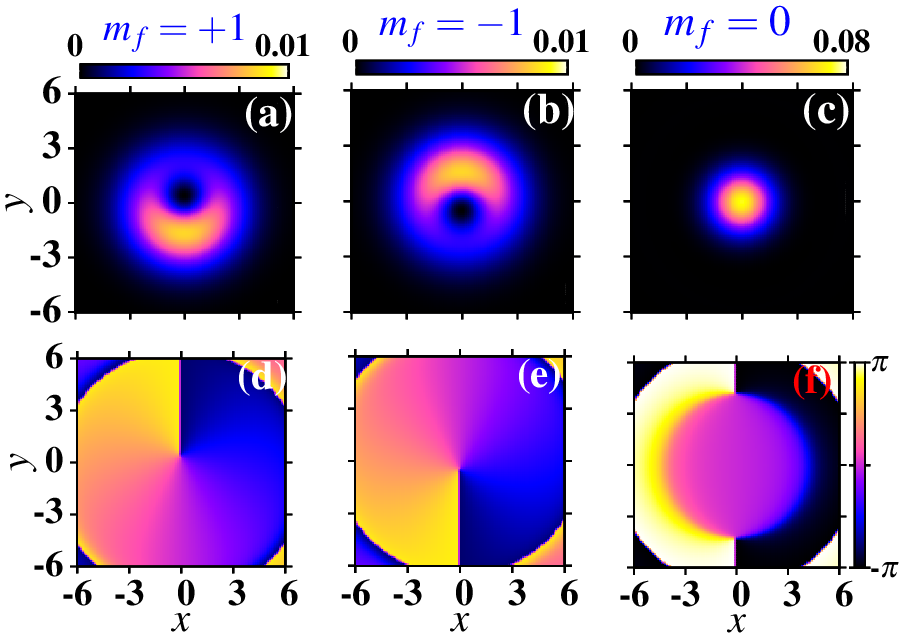}
\caption{(Color online) The 2D contour  plot of densities of the components (a) $m_f = +1$, (b)  $m_f = -1$, 
(c)  $m_f = 0$ of an  asymmetric  soliton with $c_0 = -4$ $c_1 = -0.6$  and 
$\gamma =0.5$. The corresponding phases are shown in (d) for $m_f = +1$, (e) for $m_f = -1$,
and (f)  for $m_f = 0$ components. }
\label{fig4}
\end{center}
\end{figure}

\begin{figure}[t]
\begin{center}
\includegraphics[trim = 0cm 0cm 0cm 0cm, clip,width=\linewidth,clip]{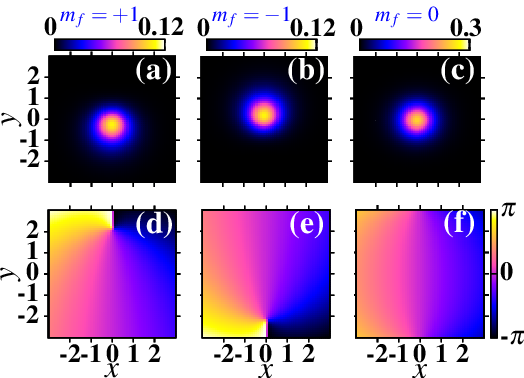}
\caption{(Color online) The 2D contour plot of densities of the components (a) $m_f = +1$, {(b)  $m_f = -1$,
(c)  $m_f = 0$} in an  asymmetric  soliton with $c_0 = -4$ $c_1 = -1.8$  and
$\gamma =0.5$. The corresponding phases are shown in (d) for $m_f = +1$, {(e) for $m_f = -1$,
and (f)  for $m_f = 0$} components.}
\label{fig3}
\end{center}
\end{figure}

 As $c_1$ is decreased further beyond $c_1^{(1)}$, e.g., for   $c_1 < c_1^{(1)}$
the axisymmetric vortex-bright solitons cease to be the ground state and  a new type of 
asymmetric soliton appears as the ground state. Nevertheless, the axisymmetric $(- 1,0,+ 1)$ 
and $(0,+ 1, + 2)$
solitons are
still  dynamically stable vortex-soliton solutions, albeit with higher energy, for $c_1^{(2)}<c_1 < c_1^{(1)}$.     
The two-dimensional contour density and phase plots of the component wave functions 
for the numerically obtained minimum-energy   asymmetric 
soliton with $c_0 = -4$, $c_1 = -0.6$ $(c_1<c_1^{(1)}) $ and $\gamma =0.5$ are shown in Fig \ref{fig4}(a)-(f).
The density corresponding to the component $\psi_0$ is axisymmetric, whereas 
the  densities corresponding to components $\psi_{\pm 1}$ are {\it asymmetric}.
 However, the total density profile (not shown here) is still radially symmetric.
The vortices in an asymmetric profile can lie along an arbitrary direction which
will be spontaneously chosen in an experiment. This is due to the fact that  Eqs. (\ref{gps-1})-(\ref{gps-2}) and
Eqs. (\ref{gpsp-1})-(\ref{gpsp-2})
are invariant under simultaneous transformations: $\phi = \tan^{-1}(y/x) \rightarrow \phi+\theta$ and 
$\psi_{m_f}(r,\phi) \rightarrow \psi_{m_f}(r,\phi+\theta) e^{ -im_f\theta}$, here $\theta$ is the angle of rotation.
Keeping $c_0$ and $\gamma$ fixed at $-4$ and $0.5$, respectively,  if we  decrease $c_1$ further from $c_1 = -0.6$,
we find that the asymmetric
ground-state  soliton continues to exist for a sufficiently large negative $c_1$ ($-0.6\ge c_1\ge -1.8$) in this case, 
beyond which it collapses. As $c_1$ is decreased from $-0.6$ to $-1.8$, the vortices in the components
$\psi_{\pm 1}$ keep on moving away from each other, 
and finally can move out of the system. This can
lead to a bright soliton with phase singularities lying at the edge of the condensate as 
is shown in Fig. \ref{fig3} for $c_0 = -4, c_1 = -1.8 $ and
$\gamma = 0.5$. In figures  Figs. \ref{fig3}(a)-(b) the solitons are bright solitons 
without any visible vortex core in density. However, the phase jump corresponding to a vortex are seen in   Figs. \ref{fig3}(d)-(e). 
It should be noted that as $c_1$ is decreased from $-0.6$ to $-1.8$, the axisymmetric
solitons shown in Figs. \ref{fig2}(a) and (c) are still stable vortex-solitons with energies higher than
the asymmetric ground state.

The  asymmetric  solitons have non-zero contribution to energy from the $c_1$ dependent terms in
contrast to the axisymmetric $(- 1,0,+ 1)$ vortex-bright soliton,
  and the details of the asymmetric soliton change as $c_1$ is changed as is illustrated
by qualitative different bright solitons in Figs. \ref{fig4} and \ref{fig3}.  
Similarly, the details of the axisymmetric $(0,+ 1, + 2)$ vortex-bright soliton are also dependent on the value of $c_1$.

\subsection{Stability of solitons}
\label{Sec-IVC}

\begin{figure}[t]
\begin{center} 
\includegraphics[trim = 2mm 1mm 2mm 2mm, clip,width=0.49\linewidth,clip]{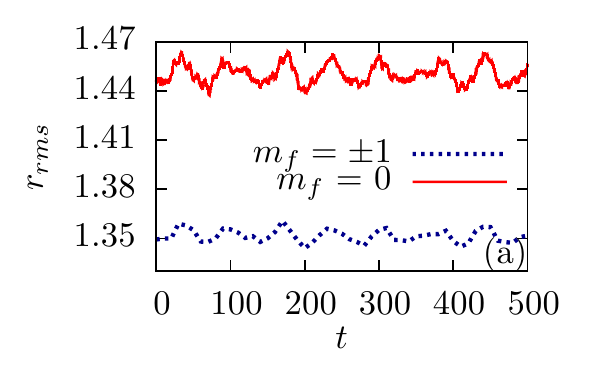}
\includegraphics[trim = 2mm 1mm 2mm 2mm, clip,width=0.49\linewidth,clip]{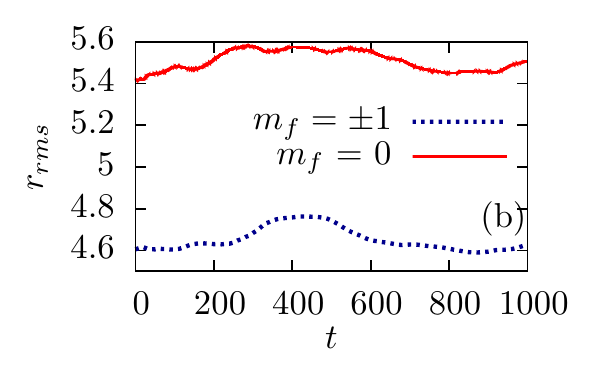}
\includegraphics[trim = 2mm 1mm 2mm 1mm, clip,width=0.49\linewidth,clip]{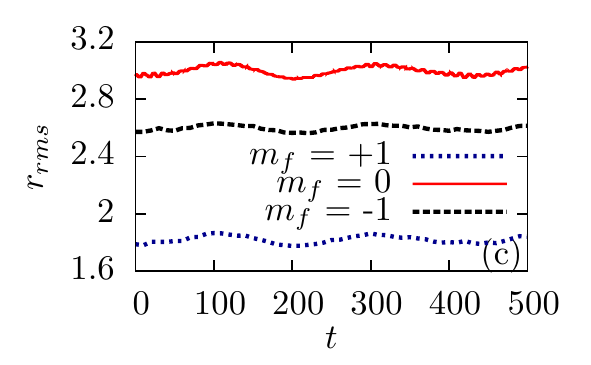}
\includegraphics[trim = 2mm 1mm 2mm 1mm, clip,width=0.49\linewidth,clip]{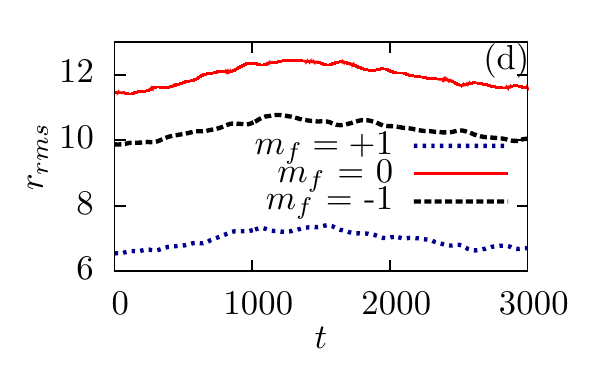}
\caption{ (Color online) Numerical result of rms  sizes of the component wave functions versus time as 
obtained in real-time simulation using the imaginary-time profiles 
of Figs. \ref{fig2}(a), (b), (c), and (d) as the initial states. 
}
\label{fig1}
\end{center}
\end{figure}

{\it Dynamical Stability:} We find that an axisymmetric $(- 1,0,+ 1)$ vector 
soliton and an asymmetric soliton can emerge as the ground states depending upon 
the choice of interaction parameters $c_0,c_1$ and $\gamma$. 
Both these solitons have zero magnetization and are dynamically stable. 
Similarly, the minimum energy axisymmetric $(0,+ 1, + 2)$ vortex-bright soliton, 
which is an excited state and which has, in general, non-zero magnetization, is dynamically stable too.

To test the dynamical stability of the $(-1,0,+1)$ vortex-bright solitons shown in 
Figs. \ref{fig2}(a) and (b) and the  $(0,+1,+2)$ vortex-bright solitons shown in 
Figs. \ref{fig2}(c) and (d), we performed real-time simulation of the imaginary-time 
profiles as the initial state over a long interval of time. The steady oscillation of the 
root mean square ($r_{rms}$) sizes of the components as shown in Figs. \ref{fig1} (a), 
(b), (c) and (d) corresponding, respectively, to solutions shown in Figs. \ref{fig2} (a), (b), (c),
and (d) demonstrates the stability of these solitons.


\subsection{Stable moving solitons} 
\label{Sec-IVD}

\begin{figure}[t]
\begin{center}
\includegraphics[trim = 0mm 0cm 0mm 0cm, clip,width=\linewidth,clip]{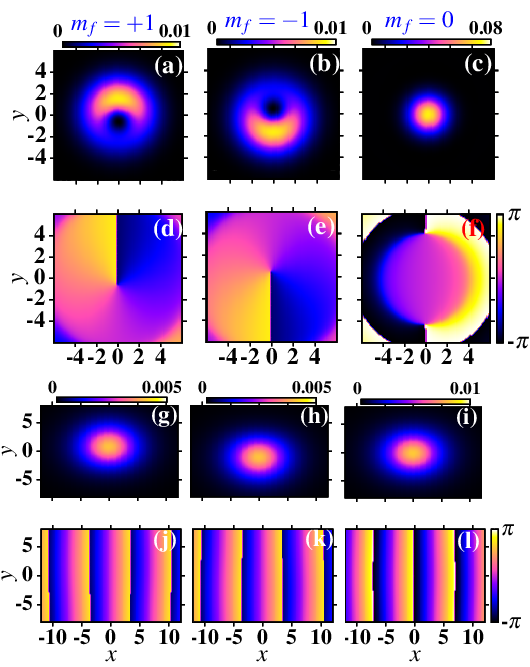}
\caption{(Color online) The 2D contour  plot of the (a) density of component $m_f=+1$, 
(b)   density of component $m_f=-1$,  (c)   density of component $m_f=0$, 
of  
the dynamically stable asymmetric soliton moving with a speed of $0.01$ along the $x$ axis for $c_0=-4, c_1=-0.25$, $\gamma=0.5$; the respective phases are shown in (d)-(f). 
   In (a) and (b) the
holes in the density profiles are antivortex and vortex, respectively.
The density and phase of the components $m_f=+1,-1$ and 0 of 
the dynamically stable soliton with the same parameters moving with 
a speed of $0.4$ along $x$ axis are shown in (g)-(i) and (j)-(l), respectively. 
}
\label{fig6}
\end{center}
\end{figure}

\begin{figure}[t]
\begin{center}
\includegraphics[trim = 0.cm 0cm 0cm 0cm, clip,width=0.47\linewidth,clip]{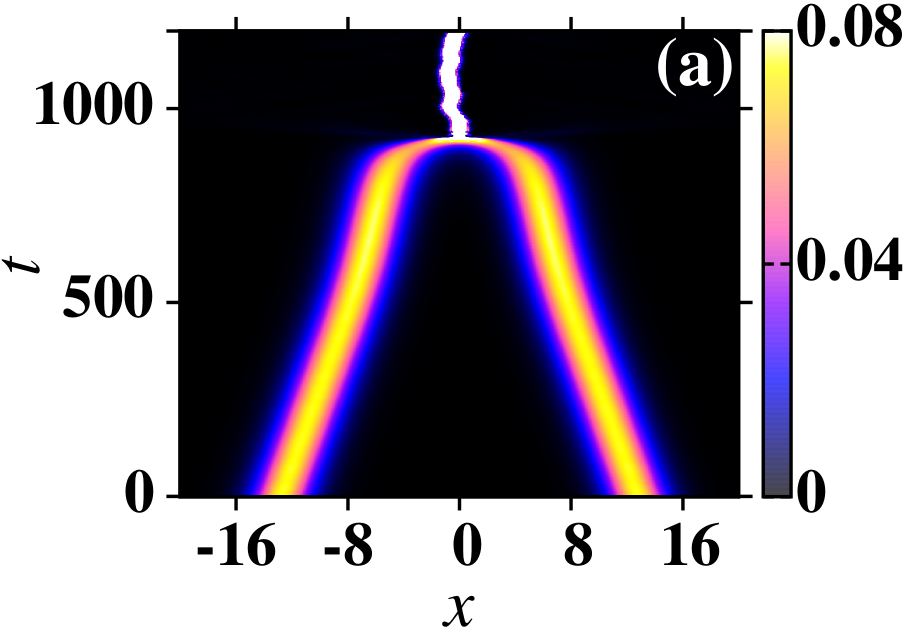}
\includegraphics[trim = 0.cm 0cm 0cm 0cm, clip,width=0.45\linewidth,clip]{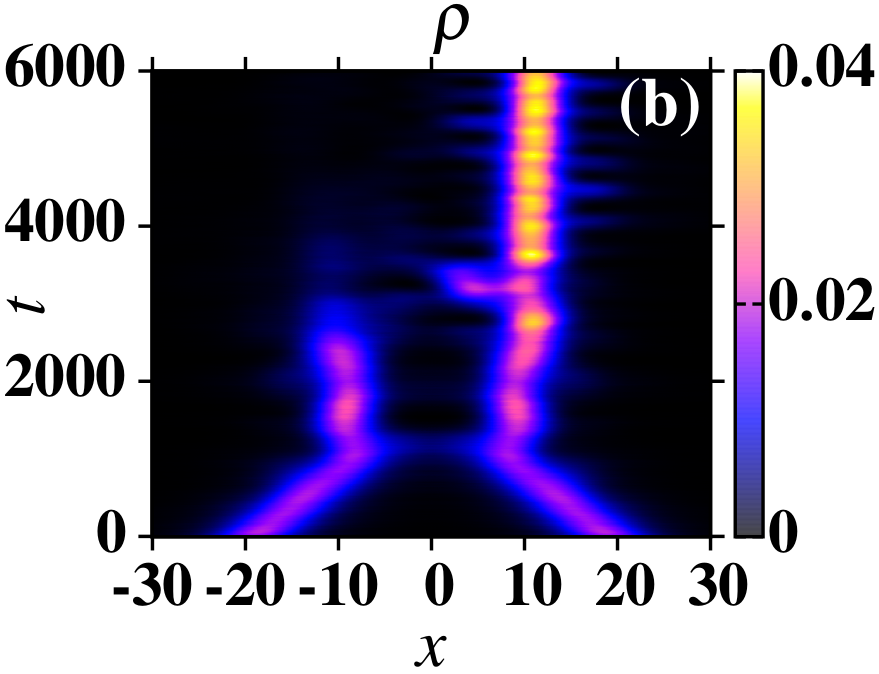}
\caption{{(Color online) (a) The 2D contour plot of the total density $\rho(x,y=0,t)$ versus $x$ and $t$
during the collision of two {\em in-phase} vortex-bright solitons, 
each with $c_0 = -4$, $c_1=-0.25$ and $\gamma = 0.5$,  moving in opposite directions along $x$ axis with 
velocity $v=0.01$. (b) shows the same for $c_0 = -2$, $c_1=-0.25$ and $\gamma = 0.5$.}
  }
\label{new-fig}
\end{center}
\end{figure}

\begin{figure}[t]
\begin{center}
\includegraphics[trim = 0.cm 0cm 0cm 0cm, clip,width=\linewidth,clip]{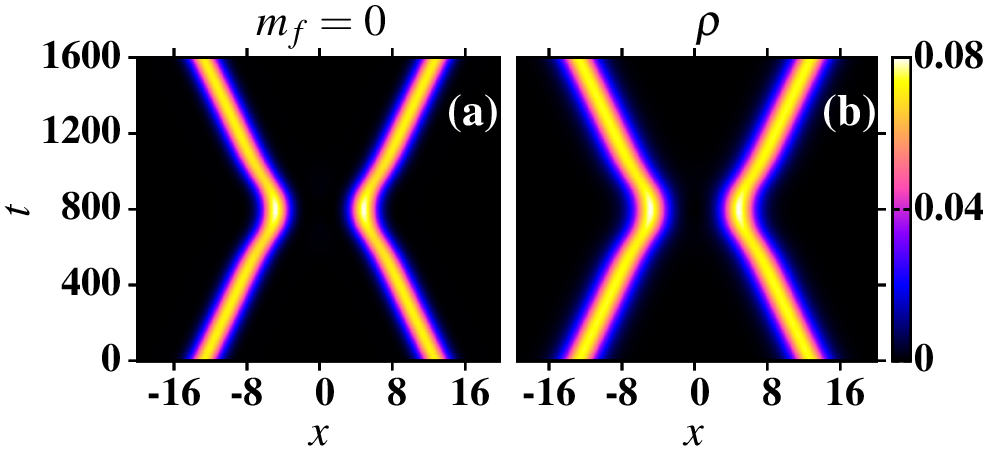}
\caption{(Color online) The 2D contour plot of (a) the  density of the $m_f =0 $ component  $\rho_0(x,y=0,t)$ and
(b) total density $\rho(x,y=0,t)$ versus $x$ and $t$ during the collision of two {{\em out-of-phase}} solitons 
considered in Fig. (\ref{fig6}) moving in opposite directions each with a speed $v=0.01$. The absence of crossing of the tracks in (a)
and (b) illustrates that the solitons rebound after the encounter.}
\label{fig7}
\end{center}
\end{figure}

\begin{figure}[t]
\begin{center}
\includegraphics[trim = 0cm 0cm 0cm 0cm, clip,width=\linewidth,clip]{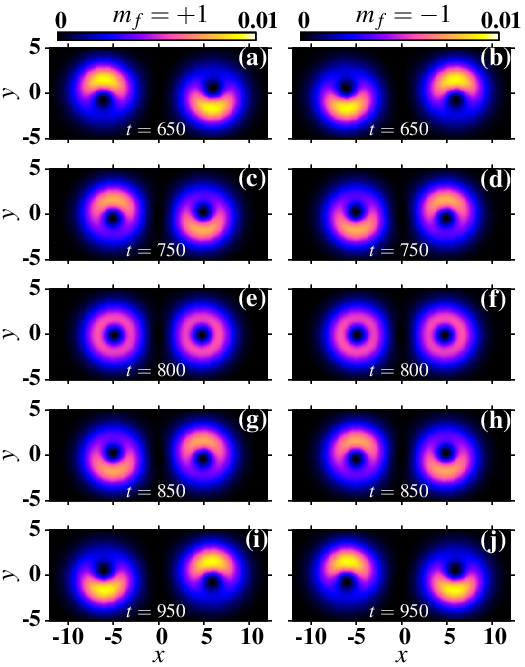}
\caption{(Color online) The  2D contour plot of densities of the $m_f=+1$ component in the
right and left moving solitons  during collision shown in Fig. \ref{fig7} at times
(a) $t=650$,  (c) $t=750$ , (e) $t=800$, (g) $t=850$, and (i) $t=950$. The same for the 
densities of the $m_f=-1$ component  in the
right and left moving soliton are shown in (b) $t=650$,  (d) $t=750$ , (f) $t=800$, (h) $t=850$, and (j) $t=950$.  At $t = 800$ the distance between the two colliding solitons 
is minimum. Holes in the density profiles of the $m_f = +1$ component correspond to antivortices,
whereas the holes in density profiles of the $m_f = -1$ component correspond to vortices.}
\label{fig8}
\end{center}
\end{figure}

In order to find the stable moving solitons, one
needs to examine the Galilean invariance of the SO-coupled Hamiltonian. 
Using Galilean transformation $x' = x - vt, y' = y, t' = t$, where $v$ is
the relative velocity along $x$ axis of the primed coordinate system with
respect to unprimed coordinate system, and using the transformation
\begin{equation}
\psi_{j}^{}(x,y,t) = \psi'_{j}(x',y',t')e^{ivx'+iv^2t'/2},\label{mov-sol}
\end{equation}
in Eqs. (\ref{gps-1})-(\ref{gps-2}), we get \cite{Gautam-3}
 \begin{eqnarray}
i \frac{\partial \psi_{\pm 1}'(\mathbf r')}{\partial t'} &=&
 {\cal H}\psi_{\pm 1}'(\mathbf r') 
\pm   c^{}_1F_z'\psi_{\pm 1}'(\mathbf r') 
+ \frac{c^{}_1}{\sqrt{2}} F_{\mp}'\psi_0'(\mathbf r')\nonumber\\
&-&\frac{i\gamma}{\sqrt{2}}\left(
  \frac{\partial\psi_0'}{\partial x'}\mp i\frac{\partial\psi_0'}{\partial y'}\right)+\frac{\gamma}{\sqrt{2}}v\psi_0',
 \label{gpsm-1}\\
i\frac{\partial \psi_0'(\mathbf r')}{\partial t'} &=&
{\cal H}\psi_0'(\mathbf r')  
+ \frac{c_1}{\sqrt 2} [F_{-}'\psi_{-1}'(\mathbf r') 
+F_{+}'\psi_{+1}'(\mathbf r')]\nonumber\\
& -&\frac{i\gamma}{\sqrt{2}}\Bigg(\frac{\partial\psi_{1}'}{\partial x} +i \frac{\partial\psi_{1}'}{\partial y'} 
  +\frac{\partial\psi_{-1}'}{\partial x'}-i\frac{\partial\psi_{-1}}{\partial y'}\Bigg)\nonumber \\
& +&\frac{\gamma}{\sqrt{2}}v(\psi_{+1}'+\psi_{-1}') 
\label{gpsm-2}.
\end{eqnarray}
Due to $v$ dependent terms in Eqs. (\ref{gpsm-1})-(\ref{gpsm-2}), the system
is not Galilean invariant. Here for the sake of simplicity, we have considered
motion along $x$ axis. In the absence of SO coupling ($\gamma = 0$), the Galilean invariance
is restored, implying that the moving solitons, {given by Eq. (\ref{mov-sol})}, can be trivially obtained by multiplying
stationary solutions of Eqs. (\ref{gps-1})-(\ref{gps-2}) with $e^{i v x}$.
This is no longer possible for $\gamma\ne 0$, in which case, the moving solitons 
are the stationary solutions, presuming that these exist, of Eqs. (\ref{gpsm-1})-(\ref{gpsm-2}) 
multiplied by $e^{i v x}$ \cite{rela,Sakaguchi,Liu}. 
The dependence of the shape of the soliton 
on its velocity is illustrated in Fig. \ref{fig6}, where we present the 2D contour plot of the 
density and phase of a soliton moving from left to right along $x$ axis with velocity $v=0.01$ and 
0.4 for the parameters  $c_0= -4, c_1 =-0.25,$ and $\gamma = 0.5$. The density and phase for $v=0.01$ 
in Figs. \ref{fig6}(a)-(f)
clearly    show the vortex and antivortex in components $m_f = \pm 1$, whereas 
in Figs. \ref{fig6}(g)-(l)
we find that the vortex and antivortex have disappeared. {These solutions, viz. $\psi_j(x,y,0)$ in Eq. (\ref{mov-sol}), 
obtained by solving Eqs. (\ref{gpsm-1})-(\ref{gpsm-2}) in imaginary-time simulation and then multiplied by $e^{i v x}$} 
are dynamically stable as confirmed in real-time simulation of these solutions 
using Eqs.  (\ref{gps-1})-(\ref{gps-2}).
   The moving soliton
has an asymmetric profile, whereas the stationary soliton for the same set of
parameters shown in Fig. \ref{fig2}(a) is axisymmetric. However, the density of the $m_f=0$ component  of the moving vortex-bright soliton shown in Figs. \ref{fig6}  is axisymmetric; the same is true about the total density. 
The result shown in Figs. \ref{fig6} is 
a manifestation of the lack of Galilean invariance in the present system which
makes the density profile of the moving soliton a function of its velocity $-$ both magnitude and
direction.
Keeping $c_0 = -4, c_1=-0.25$ and $\gamma = 0.5$ fixed, we  find that the 
Eqs. (\ref{gpsm-1})-(\ref{gpsm-2}) allow the self-trapped stationary solutions for $v\le 0.4$ along the $x$ axis; 
for $v>0.4$ along the  $x$ axis, no localized solitons can be found.
As we increase $v$ from zero, the vortices in components $m_f = +1$ and $m_f = -1$ start moving
away from each other along $y$ axis. Numerically, we also find that these vortices are located
on the line perpendicular to the direction of motion. The Eqs. (\ref{gpsm-1})-(\ref{gpsm-2}) 
are no longer invariant under transformations: $\phi = \tan^{-1}(y/x) \rightarrow \phi+\theta$ and
$\psi_{m_f}(r,\phi) \rightarrow \psi_{m_f}(r,\phi+\theta) e^{ -i m_f\theta}$ 
due to $v$ dependent terms. The orientation of the vortices in the stationary solutions of 
Eqs. (\ref{gpsm-1})-(\ref{gpsm-2}) along $y$ axis is the manifestation of the lack of this 
rotational symmetry.

\begin{figure}[t]
\begin{center}
\includegraphics[trim = 0.cm 0cm 0cm 0cm, clip,width=\linewidth,clip]{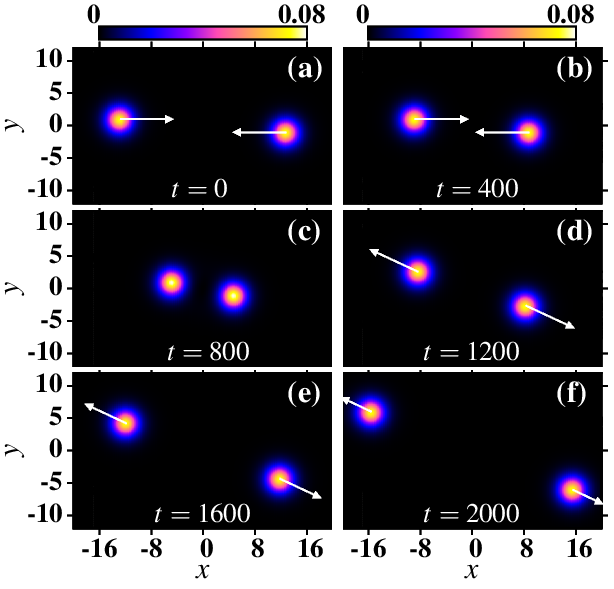}
\caption{(Color online) The  2D contour plot of total densities of the 
right and left moving {{\em out-of-phase}} solitons with the same parameters as in Fig. \ref{fig6} during collision 
with the  impact parameter $d=2$, each moving with a speed $v=0.01$, 
 at times
(a) $t=0$,  (b) $t=400$ , (c) $t=800$, (d) $t=1200$,  (e) $t=1600$, and (f) $t=2000$. 
The direction of motion of the solitons before and after collision are indicated by {white} arrows
in (a) and (d), respectively, illustrating a change in the direction of motion after collision. 
In this case the solitons repel and avoid each other without a direct encounter.} 
\label{fig9}
\end{center}
\end{figure}

The collision between two one-dimensional integrable solitons is truly elastic. 
The collision between two 2D solitons is expected to be inelastic, in general. 
We find that the two {\em in-phase} vortex-bright solitons for $c_0 = -4$, $c_1=-0.25$, $\gamma = 0.5$
and moving with speed of $v=0.01$ in opposite directions collapse after collision as is shown
Fig. \ref{new-fig}(a). In order to avert collapse, we considered the collision between the {\em in-phase} vortex-bright
solitons with $c_0 = -2$ (half of {the} previous value), $c_1=-0.25$, $\gamma = 0.5$, and $v = 0.01$.
In this case after the collision, all the atoms are transferred to one of {the} solitons as is
shown in Fig. \ref{new-fig}(b) and hence, effectively {leads} to merger as has also been observed
experimentally for scalar solitons \cite{Nguyen}. The collision in this case has similarity to the inelastic collision
between the two non-spinor bright solitons at low velocities \cite{Luis}.
However, we find that the  slowly moving  $(- 1,0,+ 1)$ solitons with asymmetric
profiles, like the ones shown in Fig. \ref{fig6}, and a phase difference of
$\pi$ can collide quasi-elastically. 
This is demonstrated by real-time simulation of two solitons,
obtained by solving Eqs. (\ref{gpsm-1})-(\ref{gpsm-2})for $c_0 = -4$, $c_1=-0.25$, $\gamma = 0.5$
and $v=0.01$ by imaginary-time propagation, 
placed initially at $t=0$ at $x=\pm 12.7$ and set into motion in opposite directions along $x$ axis with a 
speed of $v=0.01$. 
In our simulations, if we use the same
initial guess to obtain the right and left moving solitons, they end up acquiring a phase difference of $\pi$. 
 We find that the solitons come close to each other and turn back and retrace their 
trajectory without crossing each other.  This is  illustrated by the 2D contour
plot of the  axisymmetric $m_f=0$ component and total densities, (a) $\rho_0 (x,y=0,t)$ 
and (b) $\rho (x,y=0,t)$, respectively,   
versus $x$ and $t$  in Fig.  \ref{fig7}. During the collision, the asymmetric densities  of $m_f =\pm 1$ components show subtle changes
as {are} shown in Fig. \ref{fig8} through snapshots of subsequent 2D contour plots of these densities 
near the instant of closest {approach} of the two solitons. The density distribution of
$m_f =\pm 1$ components in the right and the left moving solitons are not identical as is shown in 
Figs. \ref{fig8}(a) and (b), which is again due to the break-down of the Galilean invariance.
As the left and the right moving solitons collide,
the   antivortices in the $m_f=+1$ component  in the left and the right moving solitons 
slowly move along the $y$ axis as is evident from Figs. \ref{fig8} (a), (c), (e), (g), and (i);
this is accompanied by an analogous movement of vortices in $\psi_{-1}$ as is shown in 
Figs. \ref{fig8} (b), (d), (f), (h), and (j). These changes ensure that during the course 
of the collision the two solitons 
exchange their linear momenta, and thus rebound after collision without ever crossing each other.
The repulsive collision between the two bright solitons in a quasi-1D BEC has also been 
observed experimentally \cite{Nguyen} consistent with our simulations.

\begin{figure}[t]
\begin{center}
\includegraphics[trim = 0.cm 0cm 0cm 0cm, clip,width=\linewidth,clip]{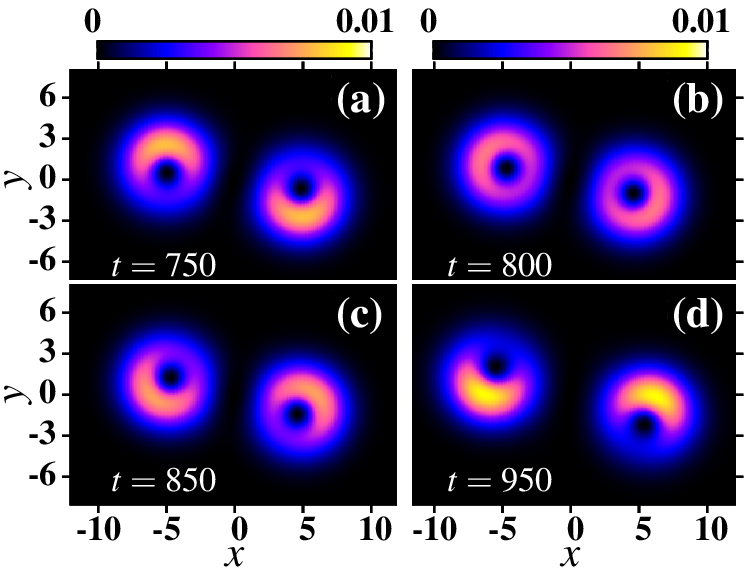}
\caption{(Color online) Dynamics of the antivortex cores,  close to the positions of closest approach, 
of $m_f = +1$ components in the right and left moving solitons for the collision
shown in Fig. \ref{fig9}. The holes in the density profiles are the antivortices.}
\label{fig12}
\end{center}
\end{figure}

We have also investigated the {\em out-of-phase} collision between two slowly  moving 
solitons along $x$ axis in opposite  directions with non-zero impact parameter $d$. 
We consider  elastic collision between two bright solitons, each with 
 $c_0 = -4$, $c_1 =-0.25$ and $\gamma = 0.5$,  placed initially ($t=0$) at 
$x=\pm 12.7, y=\pm  1$ and set into motion along the $x$ axis in opposite directions with 
speed  $v=0.01$. This collision is illustrated in Fig. \ref{fig9} by the 2D contour plot of the 
total densities of the two colliding solitons.  As in an elastic collision with non-zero 
impact parameter between two classical objects, the two solitons are deflected from their 
original trajectory conserving momentum; they do not retrace their trajectories after collision 
as in Fig. \ref{fig7}. 
The direction of motion of the solitons before and after collision are shown by white arrows in 
Figs. \ref{fig9}(a) and (d), respectively. As in the case of head-on collision shown in Fig. \ref{fig8}, 
here too the vortices in left and right moving solitons rearrange
themselves consistent with the change in the direction of motion during the collision. The
change in the density profile of the $m_f = +1$ component during the collision  is shown
in Fig. \ref{fig12}; this is accompanied by an analogous change in the density profile of the $m_f = -1$
component (not shown here), viz. Fig. \ref{fig8}.

 The {\em out-of-phase} collision illustrated in Figs. \ref{fig9} can be
theoretically analyzed by considering the collision to be equivalent to classical  elastic collision between two identical 
rigid  circular disks of equal mass  and equal scalar velocity $v$ with a non-zero impact parameter 
$d$. If the initial velocities of the two disks are 
\begin{align}
{\bf v_j} &= v \cos \theta_j \hat x+ v \sin \theta_j \hat y, 
\end{align}
 where $j=1,2$ denote the index of the disk,  $\hat x$ and $\hat y$ are the unit vectors along $x$ and $y$ axes, respectively;
then the velocity components, ($v_{jx}',v_{jy}'$), after collision are \cite{Becker}
\begin{align}
v_{jx}' &= v \cos(\theta_{3-j}-\phi)\cos \phi-v\sin(\theta_j-\phi)\sin\phi,\label{ce1}\\
v_{jy}' &= v \cos(\theta_{3-j}-\phi)\sin \phi+v\sin(\theta_j-\phi)\cos\phi\label{ce2}. 
\end{align}
Here $\phi$ is the collision angle and is related to the coordinates of
the centers of two disks at the instant of closest approach, denoted by $(C_{jx},C_{jy})$,
as 
\begin{align}
\phi = \tan^{-1}\left(\frac{C_{1y} - C_{2y}}{C_{1x}-C_{2x}}\right).
\end{align}
In our case, when the two solitons are moving along $y = +d/2$, $\theta_1 = 0$ and $y = -d/2$ and $\theta_2 = \pi$,  
$\phi$ can be written in terms of the impenetrable radius $R$ of the soliton
\begin{equation}
\phi = -\tan^{-1}\left( \frac{d}{\sqrt{4R^2-d^2}}\right),\label{ce3}
\end{equation}
where impenetrable radius can be defined as the half of the distance between the centers
of the two solitons at the distance of closest approach and is equal to $4.7$ in the present
case. Besides the collision shown in Fig. \ref{fig9} with impact parameter  $d = 2$, we also studied the
collisions with $d = 1,3,4$. Using  $\theta_1 = 0$, $\theta_2 = \pi$ in Eqs. 
(\ref{ce1})-(\ref{ce2}), we find that the magnitude of the velocities of the solitons remain 
unchanged after collision which is consistent with numerical findings. The final angles are given by $\theta_1 '= \pi+2\phi$ and $\theta_2'=2\pi+2\phi $. 
These analytic classical results for the elastic collision between two disks are 
in good agreement with the numerical result of elastic collision between two quantum 2D BEC solitons for   
$v_1 = v_2 = 0.01$, $\theta_1 = 0$, $\theta_2 = \pi$ and $R = 4.7$. 
The numerical and analytic results for $\theta_j'$'s in {\em degrees} for different values of
impact parameters are summarized in Table I.

\begin{table}
\caption{Numerical and analytical result for  the angles of the emerging solitons after collision shown in Fig. \ref{fig9}. }
\begin{center}
    \begin{tabular}{| c | c | c | c |c|}
    \hline
    \multicolumn{1}{ |c| }{}&\multicolumn{2}{ |c| }{numerical}&\multicolumn{2}{ |c| }{analytical} \\
    \hline
   $ d$ & $\theta_{1}'$ & $\theta_{2}'$ & $\theta_{1}' $ & $\theta_{2}'$\\ \hline
    1 &168.5& 348.5 &167.8 &347.8\\ \hline
    2 &155.2& 335.2 &155.4 &335.4\\ \hline
    3 &142.2& 322.2 &142.8 &322.8\\ \hline
    4 &130.2& 310.2 &129.6 &309.6\\ \hline
    \end{tabular}
\end{center}
\end{table}

\begin{figure}[t]
\begin{center}
\includegraphics[trim = 0.cm 0cm 0cm 0cm, clip,width=\linewidth,clip]{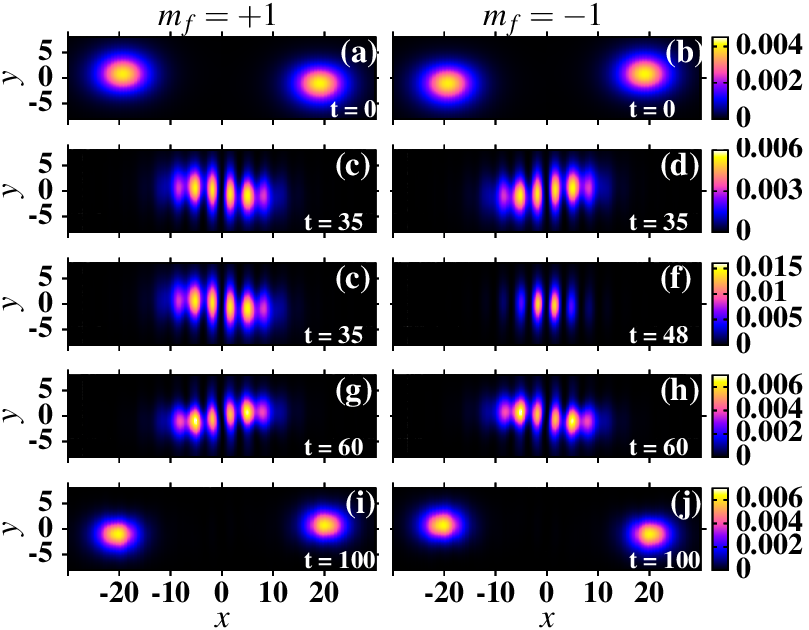}
\caption{(Color online) The 2D contour plot of densities of the $m_f=+1$ components of two {\em in-phase} vortex-bright solitons each with  
$c_0 = -4$, $c_1=-0.25$ and $\gamma = 0.5$ moving in opposite directions along $x$ axis with 
velocity $v=0.4$ at times $t=$ (a) 0, (c) 35, (e) 48, (g) 60, and (i) 100. The same for the 
$ m_f=-1$ components are, respectively, presented in (b), (d), (f), (h), and (j). 
  }
\label{fig10}
\end{center}
\end{figure}

In contrast to slowly moving vortex-bright solitons, two fast moving vortex solitons can 
pass through each other during collision irrespective of phase difference. 
To demonstrate this, we consider  the head-on collision between
two {\em in-phase} vortex-bright solitons, each with $c_0 = -4$, $c_1=-0.25$ and 
$\gamma = 0.5$, moving with a speed of $0.4$ in opposite directions along the $x$ axis.
The collision is illustrated by successive snapshots of 2D contour plots {of $m_f=\pm 1$ components} in 
Figs. \ref{fig10}. In this case, the collision dynamics of two {\em out-of-phase}  
vortex-bright solitons with same $c_0,c_1,\gamma$ and $v$ has little difference
from dynamics shown in Figs. \ref{fig10}.
This figure is qualitatively different from the dynamics shown in Fig. \ref{new-fig}(b) where 
at low velocities the solitons do not pass through each other. 
However,  Figs.  \ref{fig10} reveal that at high velocities the solitons superpose and cross 
each other like normal BEC solitons in 1D \cite{1Dcol} and 2D \cite{2Dcol}. 

\section{Summary}

We have studied the formation and dynamics of  2D vortex-bright solitons in a three-component SO-coupled 
spin-1 spinor condensate using numerical solution and variational approximation of the  mean-field 
GP equation. The ground state  vortex-bright solitons are axisymmetric in the 2D plane in  the polar ($c_1>0$) and weakly
ferromagnetic ($0>c_1>c_1^{(1)}$) domains, whereas they are asymmetric in the strongly 
ferromagnetic domain     ($c_1^{(1)}>c_1>c_1^{(2)}$). For very strong ferromagnetic interaction ($c_1< c_1^{(2)}$)
the system collapsed and no solitons can be found.  
In this problem the {coupled GP equations are} not Galilean invariant.
Consequently, to obtain the dynamically stable moving solitons, {the Galilean-transformed coupled GP equations} 
have been used. The profile of the moving soliton is dependent on its velocity vector.  
In the study of collision of two moving vortex-bright solitons at small velocities, we find that
the {\em in-phase} solitons either collapse or merge into single entity, whereas out-of-phase 
solitons repel and avoid each other without ever having an overlapping profile. 
The collision between the {\em in-phase} vortex-bright solitons is thus qualitatively similar to 
the collision of two normal (non-spinor) BEC solitons in 1D  \cite{1Dcol} and 2D \cite{2Dcol}. 
In the collision of two solitons at large velocities, they form an overlapping profile 
during interaction and cross each other. Here, the phase difference between the two
solitons has little effect on the collision dynamics as the kinetic energy of the solitons
is more than sufficient to overcome any repulsion arising due to the phase difference.

\label{Sec-V}


\begin{acknowledgements}
This work is financed by the Funda\c c\~ao de Amparo \`a Pesquisa do Estado de 
S\~ao Paulo (Brazil) under Contract Nos. 2013/07213-0, 2012/00451-0 and also by 
the Conselho Nacional de Desenvolvimento Cient\'ifico e Tecnol\'ogico (Brazil).
\end{acknowledgements}


\begin{thebibliography}{99}
\bibitem{Kivshar}
 Y. S. Kivshar and B. A. Malomed, 
 Rev. Mod. Phys. {\bf 61}, 763 (1989);
F. K. Abdullaev, A. Gammal, A. M. Kam-
chatnov, and L. Tomio, Int. J. Mod. Phys. B {\bf 19}, 3415
(2005).
\bibitem{Inouye}S. Inouye, M. R. Andrews, J. Stenger, H.-J. Miesner, D. M. Stamper-Kurn, and  W. Ketterle, Nature (London) 392, 151 (1998).
\bibitem{li}
 K. E. Strecker, G. B. Partridge, A. G. Truscott, and R. G.
Hulet, Nature (London) 417, 150 (2002); L. Khaykovich,
F. Schreck, G. Ferrari, T. Bourdel, J. Cubizolles, L. D.
Carr, Y. Castin, and C. Salomon, Science {\bf 256}, 1290
(2002).
\bibitem{rb} S. L. Cornish, S. T. Thompson, and C. E. Wieman, Phys.
Rev. Lett. {\bf 96}, 170401 (2006).
\bibitem{Perez-Garcia}
 V. M. P\`{e}rez-Gar\`{c}ia and J. B. Beitia, 
 Phys. Rev. A {\bf 72}, 033620 (2005);
 S. K. Adhikari, Phys. Lett. A {\bf 346}, 179 (2005); \pra {\bf 72}, 053608 (2005);
 L. Salasnich and B. A. Malomed, Phys. Rev. A {\bf 74}, 053610
(2006).



\bibitem{Ieda}
 J. Ieda, T. Miyakawa, and M. Wadati,
 Phys. Rev. Lett. {\bf 93}, 194102 (2004);
 L. Li, Z. Li, B. A. Malomed, D. Mihalache, and W. M. Liu,
 Phys. Rev. A {\bf 72}, 033611  (2005);
 W. Zhang, \"O. E. M\"ustecaplio\v{g}lu, and L. You,
 Phys. Rev. A {\bf 75}, 043601 (2007);
 B. J. D\k{a}browska-W\"{u}ster, E. A. Ostrovskaya, T. J. Alexander, and Y. S. Kivshar,
 Phys. Rev. A {\bf 75}, 023617 (2007);
 E. V. Doktorov, J. Wang, and J. Yang,
 Phys. Rev. A {\bf 77}, 043617 (2008);
 B. Xiong and J. Gong;
 Phys. Rev. A {\bf 81}, 033618 (2010);
 P. Szankowski, M. Trippenbach, E. Infeld, and G. Rowlands,
 Phys. Rev. Lett. {\bf 105}, 125302 (2010).


\bibitem{stringari} Y. Li, Giovanni I. Martone, and S. Stringari, 
{{\em Annual Review of Cold Atoms and Molecules,} Vol. 3, (World Scientific, 2015), 201-250;} 
 V. Galitski and I. B. Spielman, Nature {\bf 494}, 49 (2013).

\bibitem{Dalibard}
 K. Osterloh, M. Baig, L. Santos, P. Zoller, and 
 M. Lewenstein, Phys. Rev. Lett. {\bf 95}, 010403 (2005); 
 J. Ruseckas, G. Juzeli\=unas, P. \"Ohberg, and M. Fleischhauer, 
 Phys. Rev. Lett. {\bf 95}, 010404 (2005);
 G. Juzeli\=unas, J. Ruseckas, and J. Dalibard, 
 Phys. Rev. A {\bf 81}, 053403 (2010);
 Z. Lan and P. \"Ohberg, 
 Rev. Mod. Phys. {\bf 83}, 1523 (2011).



 
\bibitem{Rashba}
 Y. A. Bychkov E. I. Rashba,
 J. Phys. C {\bf 17}, 6039 (1984).
\bibitem{Dresselhaus}
 G. Dresselhaus,
 Phys. Rev. {\bf 100}, 580 (1955).
\bibitem{Lin}
 Y.-J. Lin , K. Jim\'{e}nez-Garc\'{i}a, and I. B. Spielman,
 Nature {\bf 471}, 83 (2011).

\bibitem{Aidelsburger}
 M. Aidelsburger, M. Atala, and S. Nascimb\'ene, S. Trotzky, Y.-A. Chen, and I. Bloch, 
  Phys. Rev. Lett. 107,
 255301 (2011); Z. Fu, P. Wang, and S. Chai, L. Huang,
 and J. Zhang, Phys. Rev. A 84, 043609 (2011);
 J.-Y. Zhang, S.-C. Ji, Z. Chen, L. Zhang, Z.-D. Du,
 B. Yan, G.-S. Pan, B. Zhao, Y.-J. Deng, H. Zhai, S. Chen,
 and J.-W. Pan, Phys. Rev. Lett. 109, 115301 (2012);
 C. Qu, C. Hamner, M. Gong, C. Zhang, and P. Engels,
 Phys. Rev. A 88, 021604(R) (2013).
\bibitem{rela} Y. Xu, Y. Zhang, and B. Wu,
 Phys. Rev. A {\bf 87}, 013614 (2013).
\bibitem{Xu}
L. Salasnich
and B. A. Malomed, Phys. Rev. A {\bf 87}, 063625 (2013);
 L. Salasnich, W. B. Cardoso, and B. A. Malomed,
 Phys. Rev. A {\bf 90}, 033629 (2014);
 S. Cao, C.-J. Shan, D.-W. Zhang, X. Qin, and J. Xu,
 J. Opt. Soc. Am. B {\bf 32}, 201 (2015).
\bibitem{Sakaguchi}
 H. Sakaguchi, B. Li, and B. A. Malomed,
 Phys. Rev. E {\bf 89}, 032920 (2014);
 H. Sakaguchi and B. A. Malomed,
 Phys. Rev. E 90, 062922 (2014).
\bibitem{Liu}
 Y.-K. Liu and S.-J. Yang,
 Eur. Phys. Lett., {\bf 108}, 30004 (2014).
\bibitem{Gautam-3}
 S. Gautam and S. K. Adhikari,
 Laser Phys. Lett. {\bf 12}, 045501 (2015).
\bibitem{Gautam-4}
  S. Gautam and S. K. Adhikari,
 Phys. Rev. A {\bf 91}, 063617 (2015).
\bibitem{Salasnich}
 L. Salasnich, A. Parola, and L. Reatto, 
 Phys. Rev. A {\bf 65}, 043614 (2002).
\bibitem{Ohmi}
 T. Ohmi, and K. Machida, 
 J. Phys. Soc. Japan, {\bf 67}, 1822 (1998);
 T. L. Ho, 
 Phys. Rev. Lett. {\bf 81}, 742 (1998).

\bibitem{Mizushima}
 T. Mizushima, K. Machida, and T. Kita, Phys. Rev. Lett. {\bf 89}, 030401 (2002);
 Phys. Rev. A {\bf 66}, 053610 (2002).

\bibitem{gtm}S. Gautam and S. K. Adhikari, \pra {\bf 93}, 013630 (2016); Phys. Rev. A {\bf 92}, 023616 (2015);
Phys. Rev. A {\bf 91}, 013624 (2015); Phys. Rev. A {\bf 90}, 043619 (2014).    

\bibitem{campbell}
D. L. Campbell, R. M. Price, A. Putra, A. Valdés-Curiel, D. Trypogeorgos, and I. B. Spielman,
Nature Commun. {\bf 7,} 10897 (2016).

\bibitem{Luis}
 Luis E. Young-S, P. Muruganandam, and S. K. Adhikari,
 J. Phys. B {\bf 44}, 101001 (2011).

\bibitem{Nguyen}
J. H. V. Nguyen, P. Dyke, D. Luo, B. A. Malomed, and R. G. Hulet,
Nature Physics {\bf 10}, 918 (2014).
\bibitem{H_zhai}
 H. Zhai,
 Int. J. of Mod. Phys. B, {\bf 26}, 1230001 (2012).
\bibitem{Kawaguchi}
 Y. Kawaguchi and M. Ueda, 
 Phys. Rep. {\bf 520}, 253 (2012).


\bibitem{psanand} P. Muruganandam and S. K. Adhikari, J. Phys. B {\bf 36,} 2501 (2003).

\bibitem{Wang}
 H. Wang, J. Comput. Phys., {\bf 230}, 6155 (2011); {\bf 274}, 473
(2014).
\bibitem{Bao}
 W. Bao and F. Y. Lim, 
 Siam J. Sci. Comp. {\bf 30}, 1925 (2008); 
 F. Y. Lim and W. Bao, Phys. Rev. E {\bf 78}, 066704 (2008).



\bibitem{Muruganandam}
 P. Muruganandam and S. K. Adhikari, Comput. Phys.
 Commun. {\bf 180}, 1888 (2009); D. Vudragovi\'c, I. Vidanovi\'c,
 A. Bala\v z, P. Muruganandam, and S. K. Adhikari, 
 Comput. Phys. Commun. {\bf 183}, 2021 (2012);
L. E. Young-S., D. Vudragovic, P. Muruganandam, S. K. Adhikari, and A. Bala\v z, Comput. Phys. Commun. {\bf 204}, 209 (2016); B. Satari\'c, V. Slavni\'c, A. Beli\'c, A. Bala\v z, P. Muruganandam, and S. K. Adhikari,
 Comput. Phys. Commun. {\bf 200}, 411 (2016);
R. Kishor Kumar, L. E. Young-S., D. Vudragovi\'c, Antun Bala\v z, P. Muruganandam, S. K. Adhikari,
Comput. Phys. Commun. {\bf 195}, 117 (2015).


\bibitem{Martikainen}
 {J.-P. Martikainen, {\em Dynamics and excitations of Bose-Einstein
condensates}, Acedemic dissertation, Helsinki Institute of Physics, University of Helsinki (2001).}
 
\bibitem{mh} N. D. Mermin and Tin-Lun Ho, Phys. Rev. Lett. {\bf 36}, 594
(1976).
\bibitem{at} P.W. Anderson and G. Toulouse, Phys. Rev. Lett. {\bf 38},
508 (1977).

\bibitem{Becker}
 Robert A. Becker, {\em Introduction to Theoretical Mechanics}, 
 McGraw-Hill Book Company, 1954.

\bibitem{1Dcol}S. K. Adhikari, New J. Phys {\bf 5}, 137 (2003). 




\bibitem{2Dcol}P. Pedri and L. Santos, Phys. Rev. Lett. {\bf 95}, 200404 (2005);
S. K. Adhikari, J. Phys. B {\bf 47}, 225304 (2014).
\end{thebibliography}
\end{document}